\documentclass[12pt]{article}

\usepackage{amsmath,amssymb,amsfonts,float,graphics,epsfig,epstopdf,color,verbatim,tabularx,bm,multirow,appendix}
\usepackage[utf8]{inputenc}
\usepackage[T1]{fontenc}
\usepackage{xcolor}
\usepackage{dsfont}
\usepackage{textcomp}
\usepackage{yfonts}
\usepackage{footnote}
\usepackage{bm}
\usepackage{subfigure}
\usepackage{mathrsfs}
\usepackage{graphicx}
\usepackage{verbatim}
\usepackage[colorlinks=true, citecolor=blue, linkcolor=blue, urlcolor=blue]{hyperref}
\usepackage{multirow}
\usepackage{tabularx}
\usepackage{diagbox}
\usepackage{colortbl}
\usepackage{braket}
\usepackage[normalem]{ulem}
\usepackage{tikz}
\usetikzlibrary{calc}
\usetikzlibrary{shapes.multipart}

\usepackage{scicite}

\usepackage{times}

\newenvironment{sciabstract}{%
\begin{quote} \bf}
{\end{quote}}

\newcommand{\comments}[1]{}

\newcommand{\stkout}[1]{\ifmmode\text{\sout{\ensuremath{#1}}}\else\sout{#1}\fi}

\makeatletter
\def\l@subsubsection#1#2{}
\makeatother

\title{Higher-Order Topological States with Cleavage-Dependent Dirac Mass}

\author%
{Hongyu Chen,$^{1,2}$  Min Long,$^{1,2}$ Chuang Chen, $^{1,2}$  Zi Yang Meng $^{1,2}$\\
\\
\normalsize $^{1}$Department of Physics and HK Institute of Quantum Science \& Technology, \\ 
\normalsize The University of Hong Kong, Hong Kong SAR, China\\
\normalsize $^{2}$State Key Laboratory of Optical Quantum Materials,  \\
\normalsize The University of Hong Kong, Pokfulam Road, Hong Kong SAR, China.\\
}

\date{}

\begin{document} 


\baselineskip24pt


\maketitle


\begin{sciabstract}
Topological quantum chemistry based on local charge profiles lacks predictive power for the crystalline cleavage of higher-order topological insulators (HOTIs). By cleaving an obstructed atomic insulator, we discover a topological phase characterized by \(e/2\)-fractional charges localized at precisely "half" of the corners, while the remaining empty corners host complementary vacancies of interstice charge. These zero-energy charge-vacancies and topological corners form a spatially balanced geometry, separately localized at four corners. Crucially, we demonstrate that the emergence of corner zero modes dictates that specific dangling bonds\text{—}acting as the mass of a Dirac fermion\text{—}must explicitly expose in, and subtly slope toward, the corner regions. This strict directionality is verified by the anisotropic evolution of the mass term within a \((2+1)\)-dimensional parameter space. Moreover, we find that the topological corners acquire lower entanglement entropy compared to the bulk, a behavior opposite to that of the real-space energy distribution which forms an energy-entropy compensation, essentially derived from the topological charge compensation. Our work paves the way for the local chemical environment at topological boundaries, and demonstrates the higher-order quantum-transport counterparts for high-energy Dirac physics.
\end{sciabstract}


\section*{Introduction}
The $k$-dimensional conducting boundaries of $d$-dimensional higher-order topological insulators (HOTIs) with open boundaries along more than one direction ($d-k$ $\textgreater$ 1) are not easy to classify in a unified manner, as the Berry phase cannot continuously evolve at the ($d-k$ = 1) boundary \cite{Xiao2010}. At second-order boundaries ($d=2$, $k=0$), $e/2$, $e/3$, $e/4$, $3e/4$ and $e/6$ charged corners are protected by $C_3$, $C_4$, $C_6$ rotation, mirror, reflection and the combined rotation-inversion ($S_4$) symmetry in two-dimensional qradrupole topological insulators \cite{Benalcazar2017,Song2017,Benalcazar20172,Ezawa2018,Park2019,Schindler2018,Mao2022,Benalcazar2019,Takahashi2021,Chen2021,Queiroz2019,Zhang2020,Lin2024,Langbehn2017,Han2024,zhangDiscrete2025}. 
Later on, third-order corner states in the octupole topological insulator ($d=3$, $k=0$) \cite{Benalcazar20172}, hinge states ($d=3$, $k=1$) protected by $C_{3z}$ symmetry \cite{Zhang2020} and the combined rotation-time-reversal ($C_{4z}\mathcal{T}$) symmetry \cite{Schindler2018}, are demonstrated in three-dimensional crystals.
With the topological higher-order boundaries, interface superconductor \cite{Langbehn2017}, quantized circular dichroism of chiral hinge states \cite{Pozo2019}, axion insulator \cite{Xu2019}, Majorana Kramers pairs \cite{Hsu2018}, non-Abelian braiding of Dirac fermionic modes \cite{Wu2020} and massive Dirac fermion \cite{May2022} are proposed. These excitations and responses suggest the exotic physics could emerge from the topological higher-order boundaries.

Recently, it was reported that not only the symmetry but also the thickness of the step edges distinguishes the unilateral conducting edges in a three-dimensional crystal \cite{Hossain2024}. In practice, the symmetry of the superlattice, $G_{\mathrm{Superlattice}}$, is reduced to an arbitrary subgroup of the original total symmetry of primitive cell, which conceptually differs from that of certain momentum, $G_{\bm{k}}$, or Wyckoff positions (WPs), $G_{\rm WP}$. This brings the limit of the usage of symmetry indicator \cite{Slager2013,Tang2019,Po2017,Wang2019PRL} and topological quantum chemistry (TQC) \cite{Bradlyn2017,Song2020,Nelson2021,Hwang2025,Po2018,Elcoro2021,Wang2024} to describe the precise details of HOTI phase. For instance, TQC defines obstructed atomic insulators (OAIs), with obstructed wannierization at atomic sites, as first-order topologically trivial but possessing localized interstitial charges \cite{Bradlyn2017}. Although the interstice charges are conjectured to underlie second-order corner states \cite{Wang2022}, the topological indicator and the cleavage criterion for identifying a HOTI phase in an OAI remain poorly defined. This knowledge gap implies that better criteria are needed to address the importance of fine crystalline structure and electronic structure of the conducting higher-order boundaries.

\subsection*{Model and Main Results}
 In this work, we address this open question by studying the HOTI phase in an OAI \cite{Wang2024} with the tight-binding Hamiltonian with the strength of the nearest-neighboring (NN) hopping $\braket{i,j}$, the next-nearest neighboring hopping $\braket{\braket{i,j}}$ and the third-neighboring hopping $\braket{\braket{\braket{i,j}}}$ decaying with distance:
\begin{equation}
H = \tilde{t}\sum_{\braket{i,j}}c_i^{\dagger}c_j +  te^{i\varphi }\sum_{\braket{\braket{i,j}}}c_i^{\dagger}c_j + t^{\prime}\sum_{\braket{\braket{\braket{i,j}}}}c_i^{\dagger}c_j,
\label{eq:eq1}
\end{equation}
where $\tilde{t}$ =1, $t$= $\frac{\sqrt{2}}{4}$, $\varphi = \pm 3/4\pi$ and $t^{\prime}$ = $ -\frac{1}{4} $.  The sign of $\varphi$ indicates the $+z$/$-z$ magnetic flux in the two sub-lattice squares [see Fig.~\ref{fig:fig1} (a)]. Through fine tuning the atomic arrangement at the edges of square patterns by rotating crystalline orientation, shifting atomic position, and flipping magnetic flux, we find that there exist the topological zero-energy double-degenerate $e/2$-fractional corner at two diagonal corners of the patterns with $C_2\mathcal{T}$ or $C_2$ symmetry ($G_{\mathrm{Superlattice}}$). The remaining corners without the fractional charge on another perpendicular diagonal line are always two spacious atomic interstices, $i.e.$, the vacancies of charged WPs of $4c$ which can be visualized by the distribution of bond-center Wannier functions (WFs). Interestingly, the vacancies and the fractional charge both disappear in the topologically-trivial patterns even with higher global symmetry $G_{\mathrm{Superlattice}}$ present. Comparing this topologically trivial superlattice with higher symmetry, we further find that the specific exposed NN dangling bonds, which induce the phase transition from Dirac metal to OAI, determine the zero modes and possess anisotropy. The massive Dirac fermion as the criterion for the indicator of second-order topological boundaries against symmetry explains not only the "half" charged corners but also the requirement of directionality of the specific dangling bonds in the corner regions. Inspired by the charge compensation between the zero-energy fractional charge and the cleavage-removed charge, we calculate the real-space entanglement entropy as well as the energy distribution of the topological corners and discover the exotic conversely evolved energy and entropy during the cleavage-induced generalized thermodynamic processes in the topological corner region.   

\subsection*{Crystal Cleavages}

An OAI in Eq.~\eqref{eq:eq1} can be realized by tuning the NN hopping [See Supplementary Material (SM)~\cite{suppl} Figs.~\ref{fig:fig1} (a) and S1-S3] from the atomic limit in which the energy levels are localized similar to those of molecules to the gapped phase which possibly yields in-gap fractional corner states \cite{Wang2022}, passing through an intermediate Dirac-metal phase \cite{Wang2024}. To investigate the detailed structures of the OAI, we additionally consider the gapped phase with a flipped magnetic flux configuration in one sub-lattice [See SM~\cite{suppl} Figs. S4-S5]. For the superlattice in [1 0] $\times$ [0 1] orientation, we consider two lattice configurations centered at $C_2\mathcal{T}$ ($C_2$) center [See Figs.~\ref{fig:fig1} (b) and (f) for patterns 1 and 5] and at $C_4$ center [See Figs.~\ref{fig:fig1} (c) and (g) for patterns 2 and 6]. Then we rotate the crystal orientation counterclockwise yielding the superlattice patterns along [1 1] $\times$ [-1 1] [See Figs.~\ref{fig:fig1} (d) and (h) for patterns 3 and 7] and [1 2] $\times$ [-2 1] [See Figs.~\ref{fig:fig1} (e) and (i) for patterns 4 and 8]. We find that the fractional charge appears in the energy spectrum of patterns 1, 4, 5, 8 and the two charged corners are located at distinct sub-lattices [see Fig.~\ref{fig:fig1} (j) and SM~\cite{suppl} Fig. S2].     

\begin{table}[h]
\centering
\caption{Features of 8 superlattice patterns: (1) The symmetry group of the superlattice pattern 1(5), 2(6), 3(7), 4(8) where the updated symmetry for pattern 5, 6, 7, 8 by flipping magnetic flux is labeled in parentheses if it's different from the original one. (2) $\rm WP_{edge}$ represent exposed WPs along edges.}
\begin{tabular}{c|c|c|c|c}
Pattern& 1 (5)& 2 (6)& 3 (7)& 4 (8)\\
\hline
$G_{\mathrm{Superlattice}}$&$C_2\mathcal{T}$ ($C_2$)&$C_2$, $C_4$ &$C_2$,$C_4\mathcal{T}$ ($C_2$, $C_4$)&$C_2\mathcal{T}$ ($C_2$)\\ 
$\rm WP_{edge}$&4c&2a,2b&2a,4c&2b,4c\\
\end{tabular}
\label{tab:tab1}
\end{table}
 
The summary of patterns 1-8 in accordance with the symmetry and WPs is shown in Table~\ref{tab:tab1}. We find that the fractional charge in patterns 1 and 4 coexists with the $C_2\mathcal{T}$ symmetry and the fractional charge in patterns 5 and 8 coexists with the $C_2$ symmetry. 

These results show that: ($i$) Symmetry does not always protect the HOTI phase since patterns 6 and 7 preserve $C_2$ symmetry and even higher $C_4$ symmetry,but possess no corner states. ($ii$) When projecting the eigenstates onto real space, fractional charge and the polarized edges are both localized in the atomic positions with the dangling bonds of NN hopping, $i.e.$, adjacent to the atomic interstice at WP 4$c$.  ($iii$) The cleavage where edges passing through interstice charge at WP 4$c$ cannot ensure the emergence of fractional charge but will give rise to the nonzero-energy edge polarization as shown by patterns 3 and 7 [see SM~\cite{suppl} Table S2 and Fig. S4].

\subsection*{Topological Charge Compensation}

To examine the difference between the second-order topologically trivial and nontrivial patterns, we plot the real-space distribution of charge via symmetric WFs \cite{Marzari1997,Soluyanov2011,Marzari2012,Khalaf2021,Luo2023,Xu2024,Gunawardana2024,Po20182,Song2019,Koepernik2023} in Fig.~\ref{fig:fig2}. Considering the obstructed wannierization at atomic sites, we additionally choose multiple-symmetry WPs, 2$a$, 2$b$, 4$c$ as centers and assign the amplitudes according to the adjacent bonded atoms. We design the probability of bond-type trial functions for WFs based on the averaged "amplitude distance":
\begin{equation}
P_n^{\rm Bond} =  \frac{1}{n\{\braket{i,j}\}}\sum_{\braket{i,j}} P_n^{\rm Atom} =  \frac{1}{n\{\braket{i,j}\}}\sum_{\braket{i,j}}\sum_{\alpha}|a_n^{\alpha,\bm{r}}|^2,
\end{equation}
where $a_n^{\alpha,\bm{r}}$ is the amplitude projected onto the nearest atomic positions $\bm{r}$ from the $\alpha$-th component in the eigenvector of the $n$-th Bloch band. The coordinates of atomic sites and bonds are shown in Fig.~\ref{fig:fig2} (a). $\{$$P_n^{\rm Bond=WP1}$, $P_n^{\rm Bond=WP2}$, ...$\}$ and $\{$$P_n^{\rm Atom1}$, $P_n^{\rm Atom2}$, ...$\}$ depend on the tight-binding basis. Gaussian functions with the volumes assigned by $P_n^{\rm Bond}$ or $P_n^{\rm Atom}$ localized in the unit cell both can serve as the starting guess for WFs. We first perform tests on the Hamiltonians of the periodic primitive cells [see SM~\cite{suppl} Figs. S5-S12], in which the localized feature in the atomic limit is well preserved [see SM~\cite{suppl} Figs. S5 and S9]. Then we remain this original gauge from tight-binding basis for plotting the subsequent charge of superlattice patterns. 

Strikingly, we find anomalous vacancies in the topologically nontrivial geometries when aligning the charge centers. Since the $P_n^{\rm Bond}$ are evaluated by $P_n^{\rm Atom}$, a higher $P_n^{\rm Bond}$ originates from stronger bonding with adjacent atoms and represents a more localized bonding state at the crystal interstice. In this way, the cleavage of the crystal can not only leave the dangling bonds at the boundaries but also deplete the interstice charge due to the absence of bonding atoms. In our nearest-neighbor approximation, for a high-symmetry WP, $P_n^{\rm Bond}$ = 0 when the adjacent atoms are all cut off, termed "vacancies" of interstice charge. It turns out that the vacancies at WP 4$c$ appear at the corners along one diagonal in topologically nontrivial patterns and the vacancy diagonal line is always perpendicular to the diagonal hosting the $e/2$ fractional charges [see Figs.~\ref{fig:fig2} (b-f) and Table~\ref{tab:tab2}]. 

\begin{table}[h]
\centering
\caption{(1) Zero-energy double-degenerate $e/2$-fractional corner where $\{\nwarrow, \nearrow, \searrow, \swarrow\}$ represent the positions of four corners. (2) Vacancies of the interstice charge at WP $4c$. }
\begin{tabular}{c|c|c|c|c}
Pattern& 1 (5)& 2 (6)& 3 (7)& 4 (8)\\
\hline
$\rm Fractional\,charge $&$\ket{\nwarrow \pm \searrow}$&-&-&$\ket{\nearrow\pm \swarrow}$\\
$\rm Vacancies$ &$\{\nearrow$, $\swarrow\}$&-&-&$\{\nwarrow$, $\searrow\}$\\
\end{tabular}
\label{tab:tab2}
\end{table}

We then plot the real-space total charge according to the traditional atom-center WFs and our defined bond-center WFs for superlattice patterns in Fig.~\ref{fig:fig3}. For the topologically nontrivial patterns 1 [Figs.~\ref{fig:fig3} (a) and (b)], 4 [SM~\cite{suppl} Figs. S16(b) and S16(c)], 5 [SM~\cite{suppl} Figs. S13(b) and S13(c)], 8 [SM~\cite{suppl} Figs. S16(d) and S17(e)], the corners without topological charge are large crystalline interstices and become the prominent vacancies after the charge is redistributed via bonding. Conversely, for the topologically trivial patterns 2 [Figs.~\ref{fig:fig3} (c) and (d)], 3 [SM~\cite{suppl} Figs. S15(b) and S15(c)], 6 [SM~\cite{suppl} Figs. S14(b) and S13(c)], 7 [SM~\cite{suppl} Figs. S15(d) and S17(e)], the redistributed charge spreads over the superlattice patterns. Besides, in the atom-center WFs, the charge at WP 4$c$ possesses the highest amplitude. For patterns 6 and 7 with even higher $C_4$ symmetry but without the anomalous vacancy at the corners, fractional charge disappears. This balanced geometry/neutrality between the topological fractional charge and cleavage-removed interstice charge verifies the importance of the fine electronic structure at the boundaries in addition to symmetry for HOTIs. It indicates that the redistribution of interstice charges at the boundaries can give rise to different phases in HOTIs even when the cleaved crystals look similar in terms of symmetry \cite{Hossain2024}.

\subsection*{Dirac Mass and Dangling Bonds}
Since both the topological corner states and polarized edges rely on the exposure of NN dangling bonds at WP 4$c$, we subsequently illustrate the topology here from the perspective of phase transition upon tuning the NN hopping at WP $4c$. The intermediate metallic phase between the atomic limit and the OAI contains Dirac points \cite{Wang2024} [see SM~\cite{suppl} Secs. 1 and 4, for the detailed mathematical transformation from Bloch bands to the Dirac Lagrangian] like previous 2D second-order HOTIs with $C_4$ symmetry \cite{Wen1992,Benalcazar20172,May2022}. Hence, the charge at WP $4c$ in this OAI corresponds to the mass (m) of a massive Dirac fermion in the low-energy approximation \cite{Sinha2025PRB} which is further proved by the lifted degeneracy \cite{Wu2020} of zero modes under $C_2$-breaking perturbation [see SM~\cite{suppl} Figs. S17(a-c)]. Similar to previously proposed HOTIs \cite{Benalcazar2017}, the fractional charge discovered here also comes from both the bulk electric dipole moment [see Fig.~\ref{fig:fig4} (a)] and edge polarization [see Fig.~\ref{fig:fig4} (b)]. Then the topological corners could be regarded as originating from the gapped (massive) polarized edge states \cite{ren2020,zhangDiscrete2025}. The $\mathrm{sgn}$(m) at the four edges can be obtained from rotation symmetry where edges with opposite $\mathrm{sgn}$(m) will yield $e/2$ zero modes at the connected corners \cite{May2022}. In patterns 1(5) and 4(8), where $G_{\mathrm{Superlattice}}$ is $C_2\mathcal{T}$ ($C_2$), the $C_2$ rotation enforces the opposite edges to have opposite $\mathrm{sgn}$(m): +, +, -, -, which can yield two topological corner states along one of the diagonal lines [see Fig.~\ref{fig:fig4} (c)].

For comparison, in previous $C_4$-symmetric HOTIs \cite{Wen1992,Benalcazar20172,May2022}, the $C_4$ rotation enforces that adjacent edges have opposite $\mathrm{sgn}$(m): +, -, +, -, which yields four topological corner states [see Fig.~\ref{fig:fig4} (d)]. These two different results for the superlattice in the same square shape arise from the two different lattice gauges. Since the valence bonds at WP 4$c$ induce the zero-energy gap as Dirac mass, we extracted the simplified two-band models $H(\bm{k}) = \hat{\bm{d}}_i(\bm{k})\cdot\bm{\tau}_i$ from Eq.~\eqref{eq:eq1} and the $C_4$-symmetric Hamiltonian \cite{May2022} [see SM~\cite{suppl} Sec. 5]. The spectra of thus obtained Hamiltonian are in (2 + 1)-dimensional space of $\{k_x, k_y, E(k_x, k_y)\}$, and can reflect the system's topological properties \cite{Qi2006,Wieder2018,Hwang2019,Ahn2019,Wieder2020}. By tracing the winding of $\hat{\bm{d}}$-vectors along the four edges of the Brillouin zone, we find that the winding along four edges of the Brillouin zone can be distinguished for $C_2\mathcal{T}$- and $C_4$-symmetric HOTIs as shown in Figs.~\ref{fig:fig4} (e) and (f). In $C_2\mathcal{T}$-symmetric HOTI (Fig.~\ref{fig:fig4} (e)), only the $\hat{\bm{d}}$-vectors of two edges are fully wound while the $\hat{\bm{d}}$-vectors of the other two will turn back in the middle of the edges which shows anisotropic evolution with respect to the reciprocal-spatial variables. In the $C_4$-symmetric HOTI (Fig.~\ref{fig:fig4} (f)), the winding of the $\hat{\bm{d}}$-vectors of four edges all completes a full circle.

As a result, the emergence of topological zero modes is determined by the relative orientation between the cleavage plane and the Dirac mass bonding direction. This relationship is formally captured by the connection between $G_{\mathrm{Superlattice}}$ and $G_{\mathrm{WP=4c}}$ at the corner. The winding patterns above reveal that a specific crystallographic direction—namely, the cleavage orientation—is topologically trivial. This explains the suppression of zero modes in patterns 3 and 7, where NN dangling bonds align along $\pm x$ or $\pm y$. In contrast, for nontrivial patterns 1, 4, 5, and 8, the NN dangling bonds in the corner regions always exhibit an angular tilt, which is essential for the emergence of these modes.

\subsection*{Energy and Entropy Distribution}

The dangling valence bonds as the corner modes and the spacious interstice charge vacancies form the charge compensation. Inspired by this charge "cancellation", we investigate the entropy and energy distribution at the boundaries induced by cleavage. To this end, we calculate the site-resolved binary entanglement entropy $S(\bm r)=  -\sum_{n=1}^{N_{\rm occupied}}[|a_n^{\alpha,\bm{r}}|^2 \ln(|a_n^{\alpha,\bm{r}}|^2) + (1-|a_n^{\alpha,\bm{r}}|^2)\ln(1-|a_n^{\alpha,\bm{r}}|^2)]$ and the energy $E(\bm{r}) = \sum_{n=1}^{N_{\rm occupied}}E_n\sum_\alpha^{N_{\rm orbital}} |a_n^{\alpha,\bm{r}}|^2$ over the real space to compare the results at boundaries and in the unperturbed bulk region. As shown in Fig.~\ref{fig:fig5} (a) and (b), and also SM~\cite{suppl} Fig. S19, both entropy and energy at the topological corner regions are very different from those in the bulk. The decreased entanglement entropy and the increased energy in the topological corner region compete with each other during the cleavage process and yield a compensated outcome.

\section*{Discussion}
We enrich the category of HOTIs by cleaving an OAI defined within the framework of TQC. We demonstrate that the two absent corners, for the topological fractional charge at two diagonal corners, are vacancies of the interstitial charge. In addition, we find that the appearance of zero modes relies on the exposure and orientation of dangling bonds associated with the Dirac mass in the corner regions. The fractional charge and vacancies together establish a balanced charge neutrality and geometry. The balance between the two lost charges (corresponding to the vacancies) and the emergent fractional charge indicates the nontrivial higher-order topology can be viewed as a charge compensation process, which in turn leads to the subsequent energy and entropy distribution at the higher-order boundaries. Our findings can be generalized to the models with topological criticalities and to materials defined within topological quantum chemistry or to materials with Fermi electrons near the points of topological phase transitions.




\section*{Acknowledgments}

{We thank Zhi-Da Song and Fajie Wang at Peking University, Haoran Xue at CUHK and Zhenyuan Yang at HKU for discussions. }

HYC, ML, CC and ZYM thank the HPC2021 system under the Information Technology Services \cite{HPC2021} and the Blackbody HPC system at the Department of Physics, University of Hong Kong, as well as the Beijing PARATERA Tech CO., Ltd. (URL: https://cloud.paratera.com) \cite{PARATERA2021} for providing HPC resources that have contributed to the research results reported within this paper.

\textbf{Funding:} HYC, ML, CC and ZYM are supported by the Research Grants Council (RGC) of Hong Kong (Project Nos. HKU C7037-22GF, 17302223, 17301924, 17301725), the ANR/RGC Joint Research Scheme sponsored by the RGC of Hong Kong and the French National Research Agency (Project No. A HKU703/22), and the State Key Laboratory of Optical Quantum Materials at HKU. 

\textbf{Author contributions:} HYC performed the calculations. HYC, LM and ZYM carried out the data analysis. CC provided the analysis of Wen-Zee-like term. HYC, ML and ZYM guided the project and wrote the manuscript with input from all authors.

\textbf{Competing interests:} The authors declare no competing interests.

{\textbf{Data and materials availability:} All data needed to evaluate the conclusions in the paper are present in the paper and/or the Supplementary Materials.}

\newpage

\begin{figure}[tb!]
\includegraphics[width=\columnwidth]{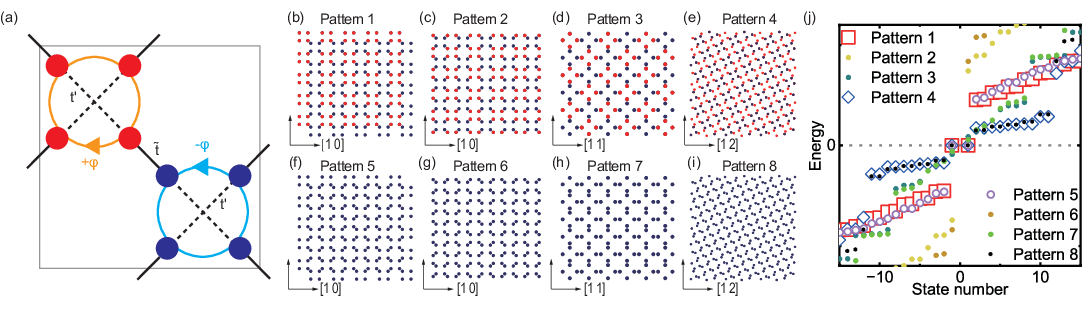}
\caption{{\bfseries Crystal cleavages with and without fractional zero modes.} Evolution of the higher-order topological phase in an obstructed atomic insulator by shifting atomic positions, rotating crystalline orientation, and flipping magnetic flux. (a) Sketch of the primitive cell with the magnetic nonsymmorphic space group of $P_c4bm$ (100.177) \cite{Aroyo2011BCS} with a magnetic/anti-translation centering operation associated with the \(C\)-face (\(P_{C}\)). The nearest-neighbor hopping $\tilde{t}$, next-nearest neighboring hopping $t$ with positive/negative phase $\varphi$ (red/blue color) and third-neighbor hopping $t^{\prime}$ are labeled. (b-e) Superlattice patterns 1-4 along [0 1]$\times$[1 0], shifted [0 1]$\times$[1 0], [-1 1]$\times$[1 1] and [-2 1]$\times$[1 2] directions. (f-i) Superlattice patterns 5-8 obtained by making all magnetic fluxes in (b–e) the same. (j) Energy levels of patterns 1-8 in which zero-energy double-degenerate corner states appear in patterns 1, 4, 5, 8.
}
\label{fig:fig1}
\end{figure}

\begin{figure*}[tp!]
\centering
\includegraphics[width=0.6\textwidth]{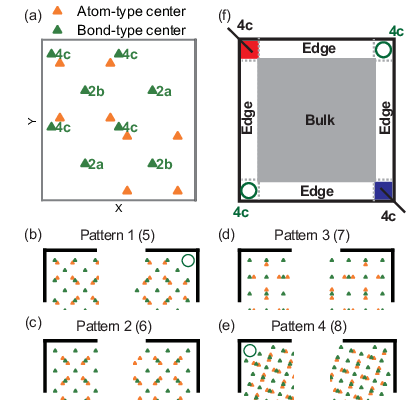}
\caption{{\bfseries Interstice charge vacancies.} Two typical Wannier functions: atom sites and bond sites. (a) Two kinds of typical positions for the gaussian trial functions at eight atom sites (orange triangles) and eight high-symmetry WPs (green triangles): $2a$, $2b$ and $4c$. Left panels in (b-e): sketch of left-upward corners for patterns 1 and 5, 2 and 6, 3 and 7, 4 and 8. Right panels in (b-e): sketches of right-upward corners for patterns 1 and 5, 2 and 6, 3 and 7, 4 and 8. The configurations for right-downward and left-downward corners are consistent with those in (b-e) by performing a $C_2$ rotation. The mismatch of periodicity between the bond-type trials and the "square-shape" atom-type trials induce the vacancies at half corners in patterns 1, 4, 5, 8, which are labeled by green circles. (f) The anomalous vacancies of WP 4$c$ and the charged corners adjacent to WP 4$c$ form a balanced geometry in the topologically nontrivial patterns, where the red (blue) region represents topological corner states at different single sub-lattice with the nearest-neighbor dangling bonds and grey region represents bulk states.
}
\label{fig:fig2}
\end{figure*}

\begin{figure*}[tb!]
\centering
\includegraphics[width=\textwidth]{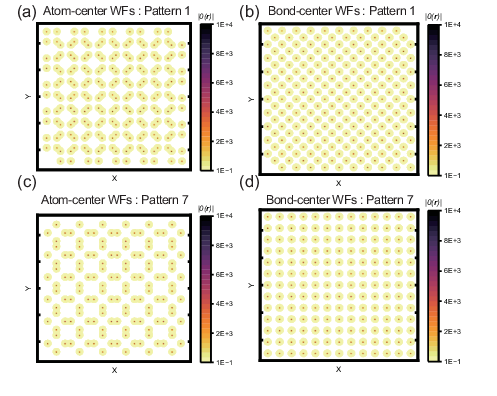}
\caption{{\bfseries Real-space charge distribution.} Comparison between the topologically nontrivial and trivial patterns in the view of real-space distribution of charge. Here, $|0(\bm{r})|$ represents the total occupied charge in the expression of central Wannier functions (WFs) for the topologically nontrivial pattern 1 with (a) atom-type and (b) bond-type centers and the topologically trivial pattern 7 with (c) atom-type and (d) bond-type centers, in which $\bm{0}$ leads to the dimensionless translation phase $e^{i\bm{k}\cdot\bm{0}}=1$.}
\label{fig:fig3}
\end{figure*}

\begin{figure}[tb!]
\centering
\includegraphics[width=0.7\columnwidth]{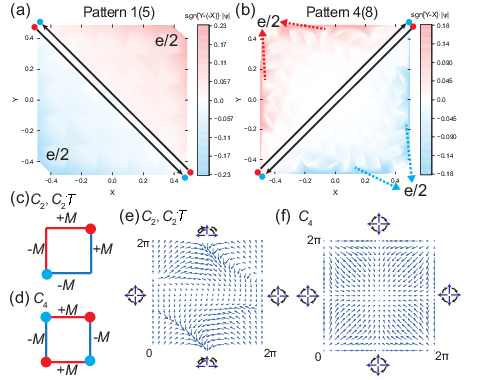}
\caption{{\bfseries Emerging higher-order topological insulators (HOTIs) with anisotropic Dirac mass vectors.} (a, b) Sketch of the 2D second-order HOTIs with the double-degenerate "half" $e/2$-charged corners which mainly arise from (a) bulk electric dipole moment for patterns 1, 5 and (b) polarized edges for patterns 4, 8. The pink and light blue gradients represent the bulk states that are nearest to zero energy. The occupied (unoccupied) corners are colored by red (blue) dots and the resultant dipoles are labeled by black arrows. (c, d) Sketch of mass of four edges for (c) a HOTI with two charged corners and (d) a HOTI with four charged corners \cite{Benalcazar20172,May2022}. (e, f)  Top view of the Skyrmion configurations, $\hat{\bm{d}}(\bm{k})$, of the dispersion relation, $H(\bm{k}) = \hat{\bm{d}}_i(\bm{k})\cdot\bm{\tau}_i$, extracted from (e) Eq.~\eqref{eq:eq1} and (f) $C_4$-symmetric Hamiltonian with four corner states \cite{Benalcazar20172,May2022} by only keeping the valence bonds as the Dirac mass which induces the zero-energy gap and neglecting the irrelevant hoppings. The four outer panels in (e) and (f) represent the winding of $\hat{\bm{d}}(\bm{k})$ along four edges of the 2D Brillouin zone.
}
\label{fig:fig4}
\end{figure}

\begin{figure}[tb!]
\includegraphics[width=\columnwidth]{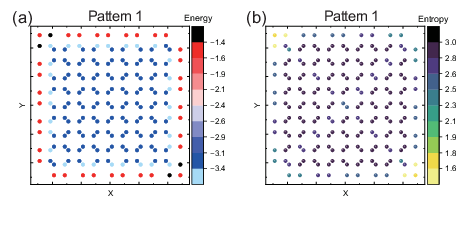}
\caption{{\bfseries Energy and entropy distribution.} Real-space distribution of the (a) the energy and (b) the entanglement entropy in the topologically nontrivial pattern 1. The energy is higher, and the entropy is lower in the topological corner region than in the bulk.
}
\label{fig:fig5}
\end{figure}


\newpage
\clearpage

\setcounter{equation}{0}
\setcounter{figure}{0}
\setcounter{table}{0}
\setcounter{page}{1}
\makeatletter
\renewcommand{\theequation}{S\arabic{equation}}
\renewcommand{\thefigure}{S\arabic{figure}}
\setcounter{secnumdepth}{3}

\section*{Supplementary materials}

\section{Bands in different phases in the model of an obstructed atomic insulator }
The Hamiltonian for the obstructed atomic insulator (OAI) \cite{Wang2024} can produce different phases by tuning the nearest-neighboring (NN) hopping $\tilde{t}$:
\begin{equation}
H = \tilde{t}\sum_{\braket{i,j}}c_i^{\dagger}c_j + te^{i\varphi }\sum_{\braket{\braket{i,j}}}c_i^{\dagger}c_j + t^{\prime}\sum_{\braket{\braket{\braket{i,j}}}}c_i^{\dagger}c_j,
\end{equation}
where $\tilde{t}$ =$\theta + \pi/2$, $t$= $\frac{\sqrt{2}}{4}$, $\varphi = \pm 3\pi/4$ and $t^{\prime}$ = $ -\frac{1}{4} $. The sign of $\varphi$ indicates the $+z$/$-z$ magnetic flux in the sub-lattice squares. 

Here we consider four periodic bulk Hamiltonians:\\
\begin{itemize}
\item Atomic limit with opposite flux in each sub-lattice: $\theta = -\pi/2$ and $\varphi = \pm 3/4\pi$;\\
\item Metallic phase with opposite flux in each sub-lattice: $\theta = 0$ and $\varphi = \pm 3/4\pi$;\\
\item Obstructed atomic insulator with opposite flux in each sub-lattice: $\theta = \pi/2$ and $\varphi = \pm 3/4\pi$;\\
\item Obstructed atomic insulator with the same flux in each sub-lattice: $\theta = \pi/2$ and all $\varphi = -3/4\pi$.\\
\end{itemize}

\begin{figure}[h]
\centering
\includegraphics[scale=0.5]{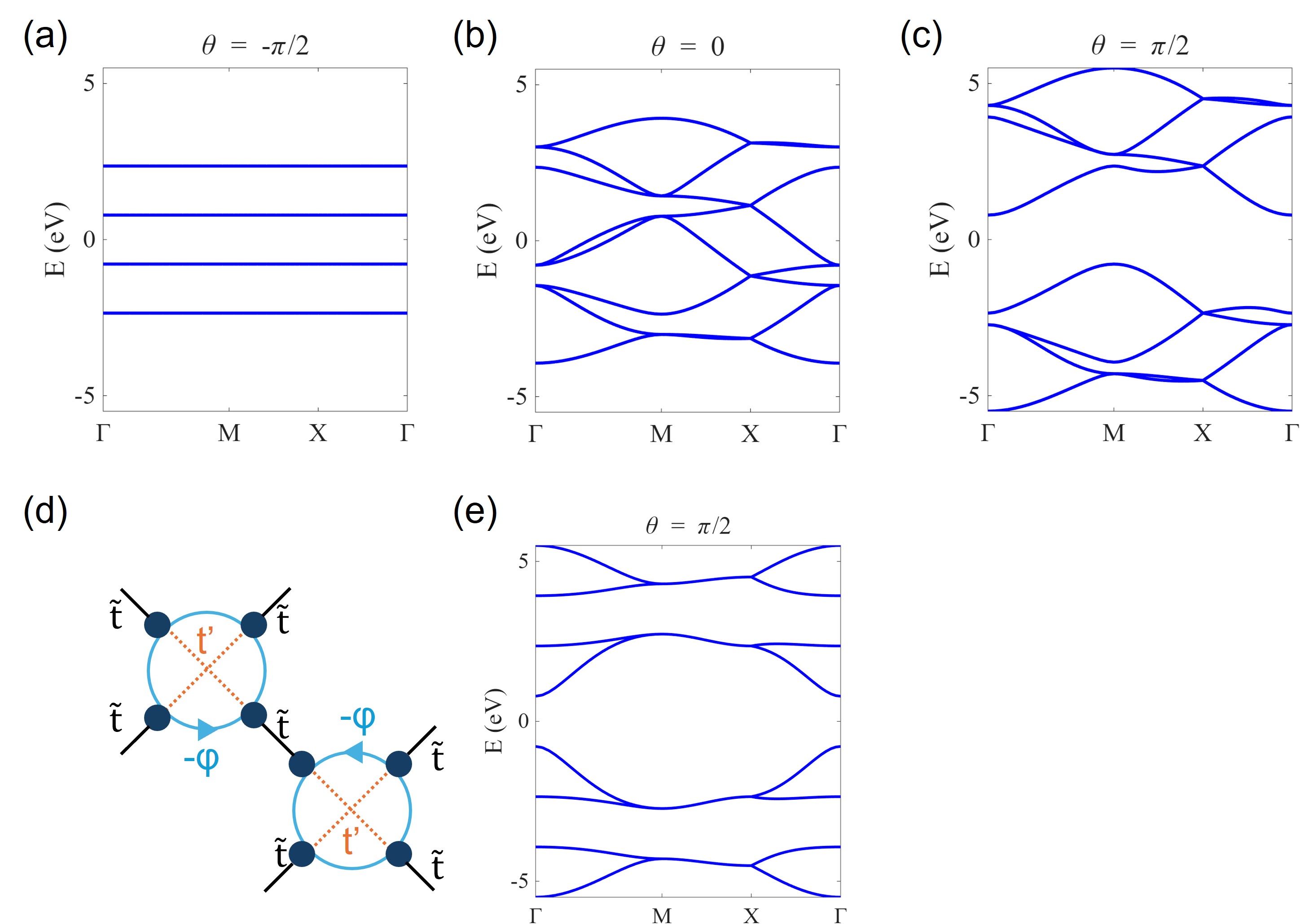}
\caption{{\bfseries Phases with respect to the electronic structures in the Manhattan lattice model in Eq. (1).} (a) Atomic limit with opposite flux in each sub-lattice. (b) Metallic phase with opposite flux in each sub-lattice in which Dirac points are along $\Gamma-X$ path. (c) Obstructed atomic insulator with opposite flux in each sub-lattice. (d, e) Obstructed atomic insulator  with the same flux in each sub-lattice.}
\end{figure}

\section{Evolving corner states by cutting style}
Eight superlattice-pattern Hamiltonians with dangling bonds:\\
\begin{itemize} 
\item Patterns 1-4 with the staggered flux: [1 0]$\times$[0 1] edge with Wyckoff-position 4$c$ dangling bonds, [1 0]$\times$[0 1] edge with Wyckoff-position 2$a$, 2$b$ dangling bonds, [1 1]$\times$[-1 1] edge and [1 2]$\times$[-2 1] edge;\\
\item Patterns 5-8 with the same flux: [1 0]$\times$[0 1] edge with Wyckoff-position 4$c$ dangling bonds, [1 0]$\times$[0 1] edge with Wyckoff-position 2$a$, 2$b$ dangling bonds, [1 1]$\times$[-1 1] edge and [1 2]$\times$[-2 1] edge.\\
\end{itemize}

\begin{figure}[h]
\includegraphics[scale=1]{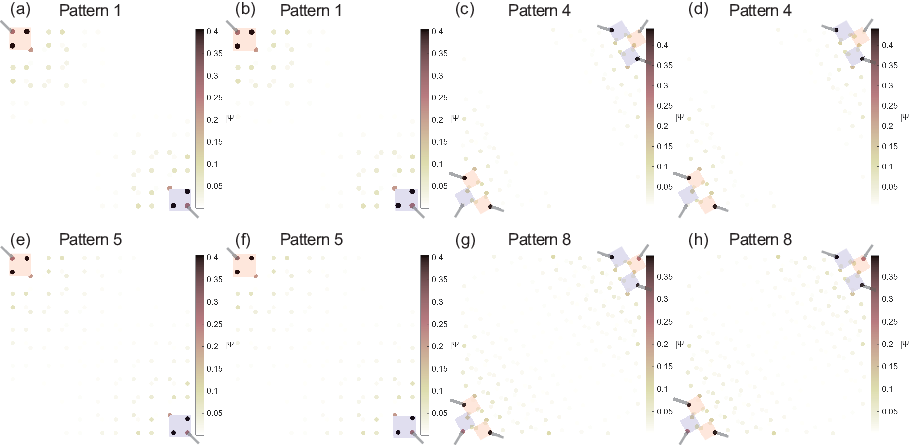}
\caption{{\bfseries Zero-energy topologically nontrivial phases}: Real-space wavefunction distribution of zero modes in (a, b) pattern 1, (c, d) pattern 4, (e, f) pattern 5, (g, h) pattern 8. The two sub-lattices with NN dangling bonds at WP 4$c$ are labeled by translucent red and blue squares with gray short lines. In patterns 1 and 5, the corner states mainly arise from the bulk moment, but in patterns 4 and 8, the corner states mainly arise from the edge polarization. }
\end{figure}

The corner charge is calculated by the electron density
\begin{equation}
\rho(\bm{r}) = Ne -e\sum_1^{N_{\rm occupied}}\sum_\alpha^{N_{\rm orbital}} |a_n^{\alpha,\bm{r}}|^2,
\end{equation}
where $Ne$ represent positive charge and $a_n^{\alpha,\bm{r}}$ represent the projected amplitude of the $\alpha$-component of $n^{th}$ wavefunction onto real-space coordinate $\bm{r}$. The charge for one corner is calculated by a two-dimensional integration over the real space:
\begin{equation}
\label{cornercharge}
Q^{\rm corner} = \sum_{(x,y) \subset \{x_{\rm corner}, y_{\rm corner}\}} \rho(x,y) dx dy.
\end{equation}
Here, to illustrate the origin of the fractional charge, we further divide the superlattice into eight sectors as shown in Fig. S3.

\begin{figure}[h]
\centering
\includegraphics[scale=0.7]{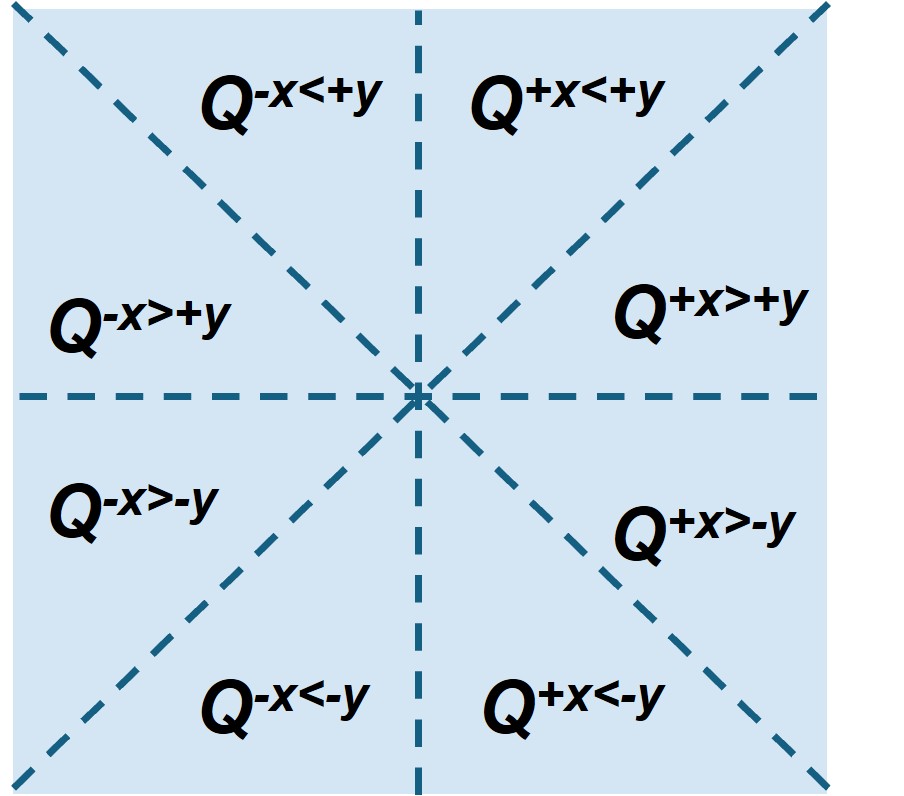}
\caption{Illustration of sections for calculating corner charge.}
\end{figure}

\begin{table}[h]
\centering
\caption{\label{t1} The calculated corner charge of topologically nontrivial patterns: 1, 4, 5, 8.}
\centering
\begin{tabular}{ccccc}\hline\hline
& Pattern 1& Pattern 4&Pattern 5&Pattern 8\\
\hline
$Q^{+x<+y}$&0&0.27e&0&0.29e\\
$Q^{+x>+y}$&0&0.23e&0&0.21e\\
$Q^{+x>-y}$&0.25e&0&0.25e&0\\
$Q^{+x<-y}$&0.25e&0&0.25e&0\\
$Q^{-x<-y}$&0&0.23e&0&0.21e\\
$Q^{-x>-y}$&0&0.27e&0&0.29e\\
$Q^{-x>+y}$&0.25e&0&0.25e&0\\
$Q^{-x<+y}$&0.25e&0&0.25e&0\\\hline\hline
\end{tabular}
\end{table}

\begin{figure}[h]
\centering
\includegraphics[scale=1]{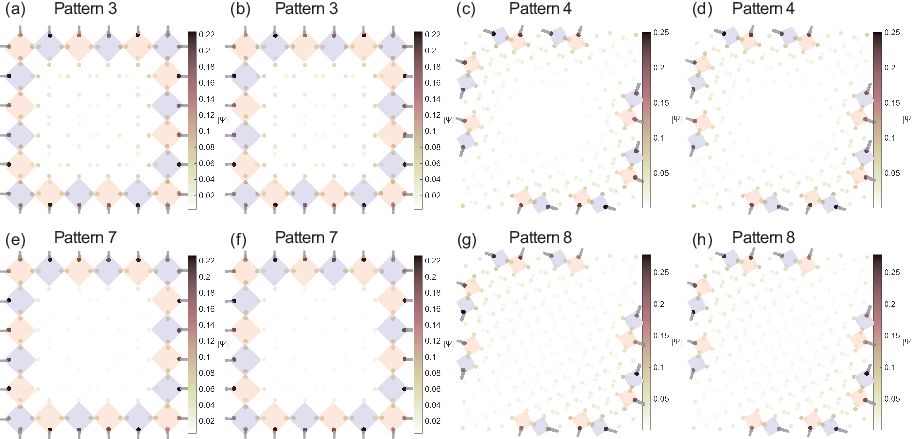}
\caption{{\bfseries In-gap nonzero-energy topologically nontrivial phases}: Real-space wavefunction distributio of nonzero-energy edge states in (a, b) pattern 3, (c, d) pattern 4, (e, f) pattern 7, (g, h) pattern 8. In patterns 3 and 7, the polarized edges do not give rise to corner states, but in patterns 4 and 8, the polarized edges give rise to the zero-energy corner states. The two sub-lattices with NN dangling bonds at WP 4$c$ and localized eigenstates are labeled by the translucent red and blue squares with gray short lines. }
\end{figure}

\begin{table}[h]
\centering
\caption{Additional features of 8 superlattice patterns: Nonzero-energy in-gap topological states (NEIGTS) where $\{\nwarrow, \nearrow, \searrow, \swarrow\}$ represent the regions of four corners, and $\square$ represent incomplete edge states which do not connect conductance bands and valence bands; gap ($\Delta$E) between the nearest bulk states above and below zero energy in superlattice patterns; distribution of the nearest bulk states above and below zero energy. }
\begin{tabular}{ccccc}\hline\hline
Patterns& 1& 2& 3& 4\\
\hline
NEIGTS&-&- & 16 $\square$ &16 $\square$ +\texttt{\tiny$\ket{\nearrow\pm \swarrow}$ + $\ket{\nearrow\pm \swarrow}$}  \\
$\Delta$ E&2.26&3.70 &4.18 &3.58 \\
Bulk state&$\{\nwarrow, \nearrow\}$ &center &$\{ \nwarrow, \nearrow, \searrow, \swarrow  \}$ & center\\\hline\hline
Patterns& 5& 6& 7& 8\\
\hline
NEIGTS&- &- &16 $\square$ &16 $\square$ +\texttt{\tiny$\ket{\nearrow\pm \swarrow}$ + $\ket{\nearrow\pm \swarrow}$} \\
$\Delta$ E&2.3 &4.20 &4.18 &4.18 \\
Bulk state&$\{\nwarrow, \nearrow\}$ &center &$\{\nwarrow, \nearrow, \searrow, \swarrow\}$ & center\\\hline\hline
\end{tabular}
\end{table}

\clearpage
\newpage
\section{Wannier functions}
We construct symmetric Wannier functions (WFs) \cite{Marzari1997,Soluyanov2011,Marzari2012,Khalaf2021,Luo2023,Xu2024,Gunawardana2024,Po2018,Song2019,Koepernik2023}, $\ket{\bm{R}_n}$, for the 12 Hamiltonians.
These Hamiltonians are in the form of tight-binding (TB) models as well as the TB basis wave functions $\chi_{p\bm{k}}$:
\begin{equation}
\begin{aligned}
&H^{TB} (\bm{k}) = \sum_{\bm{R}}t_{\bm{R}}e^{i\bm{k}\cdot\bm{R}} = \bra{\chi_{p\bm{k}}}H\ket{\chi_{p^{\prime}\bm{k}}}\\
&\chi_{p\bm{k}}(\bm{r}) = \frac{1}{\sqrt{N}}\sum_{\bm{R}}e^{i\bm{k}\cdot\bm{R}}\phi(\bm{r} - \bm{R} -\bm{o}_p).
\end{aligned}
\end{equation}
where $\bm{o}_p$ are the coordinates of atoms which obey the rule of nearest neighbors, next-nearest neighbors, and third-nearest neighbors in the formula of the TB model where we set:
\begin{equation}
\begin{aligned}
\bm{o}_{1, \,...,\, 8} = \begin{bmatrix} 1/16\\ 15/16 \end{bmatrix}, \begin{bmatrix} 7/16\\ 15/16 \end{bmatrix}, \begin{bmatrix} 1/16\\ 9/16 \end{bmatrix}, \begin{bmatrix} 7/16\\ 9/16 \end{bmatrix}, 
\begin{bmatrix} 9/16\\ 7/16 \end{bmatrix}, \begin{bmatrix} 15/16\\ 7/16 \end{bmatrix}, \begin{bmatrix} 9/16\\ 1/16 \end{bmatrix}, \begin{bmatrix} 15/16\\ 1/16 \end{bmatrix}.
\end{aligned}
\end{equation}
The $q$ Bloch bands to be analyzed can be written as:
\begin{equation}
\begin{aligned}
&\ket{\psi_{q\bm{k}}} = \sum_pC_{pq}\chi_{p\bm{k}}.
\end{aligned}
\end{equation}
We use $\delta$-form trial functions
\begin{equation}
\begin{aligned}
&\tau_{i}(\bm{r}) = \delta(\bm{r} - \bm{d}_i)\\
\end{aligned}
\end{equation}
in the home unit cell as the trial functions, $\ket{\tau_i}$ ({$n = q$} is less than or equal to the length of the Hamiltonians L: 8, 8, 8, 8, 200, 200, 144, 360, 200, 200, 144, 360). The Wannier centers (WCs), $\bm{d_n}$, are chosen at the sites to keep the total symmetry of the system in which the multiplicity of WCs in superlattice patterns is corresponding to the period of pattern of primitive cell. 

In detail, we use two kinds of $\delta$-shape trial functions in the way of multiple Gaussian functions in the home cell for considering atomic sites and bond sites for each Hamiltonian. Taking the bulk Hamiltonians as examples, the amplitudes and radius of the 8 set of Gaussian functions are $\bm{k}$-dependence. The 8-component probability of wavefunction at $\bm{k}$ for $n^{th}$ state determines the float of the amplitudes of that set of Gaussian functions where a larger probability gives a higher Gaussian function which indicates a more localized state in the real space:
\begin{itemize}
\item Atom-type $\delta$-trials:\\
\begin{equation}
\tau_{n\bm{k}}(\bm{r}) =\sum_{\alpha =1}^{N_{\rm orbital}}\frac{|a_{n\bm{k}}^{\alpha}|}{\sqrt{\pi}\sigma}e^{-\frac{|\bm{r}-\bm{o}_\alpha|}{\sqrt{2}\sigma}}
\end{equation}
where $|a_{n\bm{k}}^{\alpha}|$ is the $\alpha$-component of the $n^{th}$ eigenvector at $\bm{k}$ of the TB model, and $\sigma$ is the broadening width.
\item Bond-type $\delta$-trials:\\
\begin{equation}
\tau_{n\bm{k}}(\bm{r}) =\sum_{\alpha =1}^{N_{\rm WPs}}\frac{|A_{n\bm{k}}^{\alpha}|}{\sqrt{\pi}\sigma}e^{-\frac{|\bm{r}-\bm{w}_\alpha|}{\sqrt{2}\sigma}}
\end{equation}
\end{itemize}
where $\bm{w}_\alpha$ are the eight high-symmetry Wyckoff positions (WPs) (2$a$, 2$b$ and 4$c$) and the $\bm{k}$-dependent probabilities $A_{n\bm{k}}^{\alpha}$ at the eight WPs are the averages of the probabilities $|a_{n\bm{k}}^{\alpha}|$ of the nearest-neighboring atomic sites. Both typical trials at each $\bm{k}$ are normalized: 
\begin{equation}
\begin{aligned}
&\int {\tau^{\ast}_{n\bm{k}}}(\bm{r})\tau_{n\bm{k}}(\bm{r})d\bm{r} = 1.
\end{aligned}
\end{equation}

By these localized trial functions, we can project them onto the $q$-subspace to get the periodic Bloch-like states:
\begin{equation}
\begin{aligned}
&\ket{\gamma_{i\bm{k}}} = \sum^{q}_{1} \ket{\psi_{q\bm{k}}}\braket{\psi_{q\bm{k}}|\tau_i} \equiv u_i(\bm{r})e^{i\bm{k}\cdot\bm{r}}.
\end{aligned}
\end{equation}
The constructed WFs can be obtained by Fourier transformation:
\begin{equation}
\begin{aligned}
&\ket{\bm{R}n} = \frac{1}{N_k}\sum^{N_k}_1e^{-i\bm{k}\cdot\bm{R}}\ket{\gamma_{n\bm{k}}},\\
\end{aligned}
\end{equation}
where $R$ indicates the index for periodic primitive cells, so the central WFs at $R=0$ (where $e^{-i\bm{k}\cdot\bm{R}}$=1) can be obtained directly from the periodic part of Bloch state at home cell 
\begin{equation}
\begin{aligned}
&\bm{0}_n(\bm{r}) = \frac{1}{N_k}\sum^{N_k}_1e^{i\bm{k}\cdot\bm{r}}\tau_{n\bm{k}}(\bm{r}).
\end{aligned}
\end{equation}
To maintain the atomic-limit feature of each isolated band as shown in Figs. S5 and S9, we do not perform the additional orthogonalization for $\ket{\gamma_{i\bm{k}}}$. The localization of WFs is still checked by overlap matrix
\begin{equation}
S_{mn}(\bm{k}) = \braket{\gamma_{m\bm{k}}|\gamma_{n\bm{k}}} \neq 0.
\end{equation}
For superlattice Hamiltonian, there is no Brillouin zone for the disks so that the $\delta$-trials are not $k-$dependence. We just sum all the occupied states to evaluate the real-space distribution of total charge:
\begin{equation}
\begin{aligned}
\bm{0}(\bm{r}) &= \sum^{N_{\rm occupied}}_n \frac{1}{N_k}\sum^{N_k}_1e^{i\bm{k}\cdot\bm{r}}\tau_{n}(\bm{r})\\& = \sum^{N_{\rm occupied}}_n\tau_{n}(\bm{r}),
\end{aligned}
\end{equation}
as shown in Figs. S13-S16. 

\begin{figure}[h]
\centering
\includegraphics[scale=0.28]{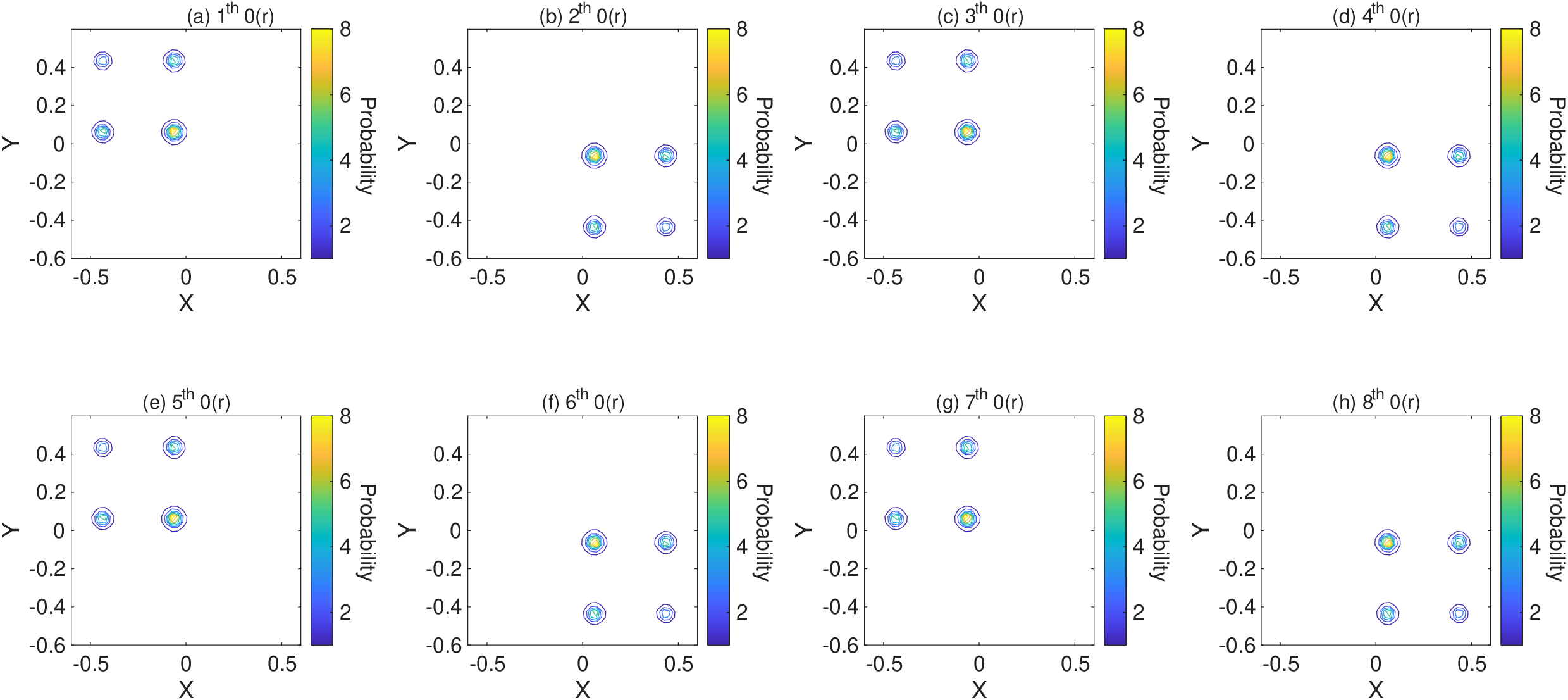}
\caption{{\bfseries Atom-type Wannier functions for the atomic-limit phase in Figure. S1(a).}}
\end{figure}

\begin{figure}[h]
\centering
\includegraphics[scale=0.28]{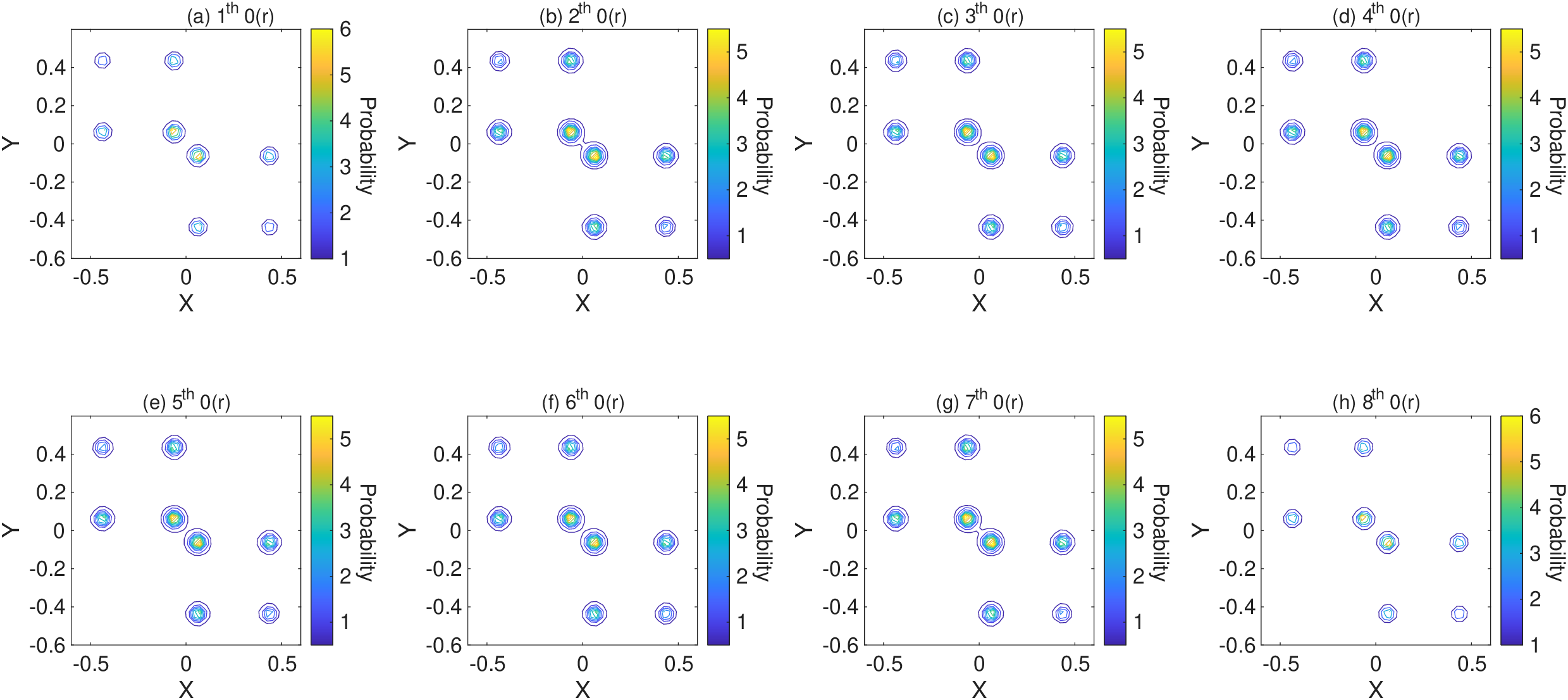}
\caption{{\bfseries Atom-type Wannier functions for the metallic phase in Figure. S1(b).}}
\end{figure}

\begin{figure}[h]
\centering
\includegraphics[scale=0.28]{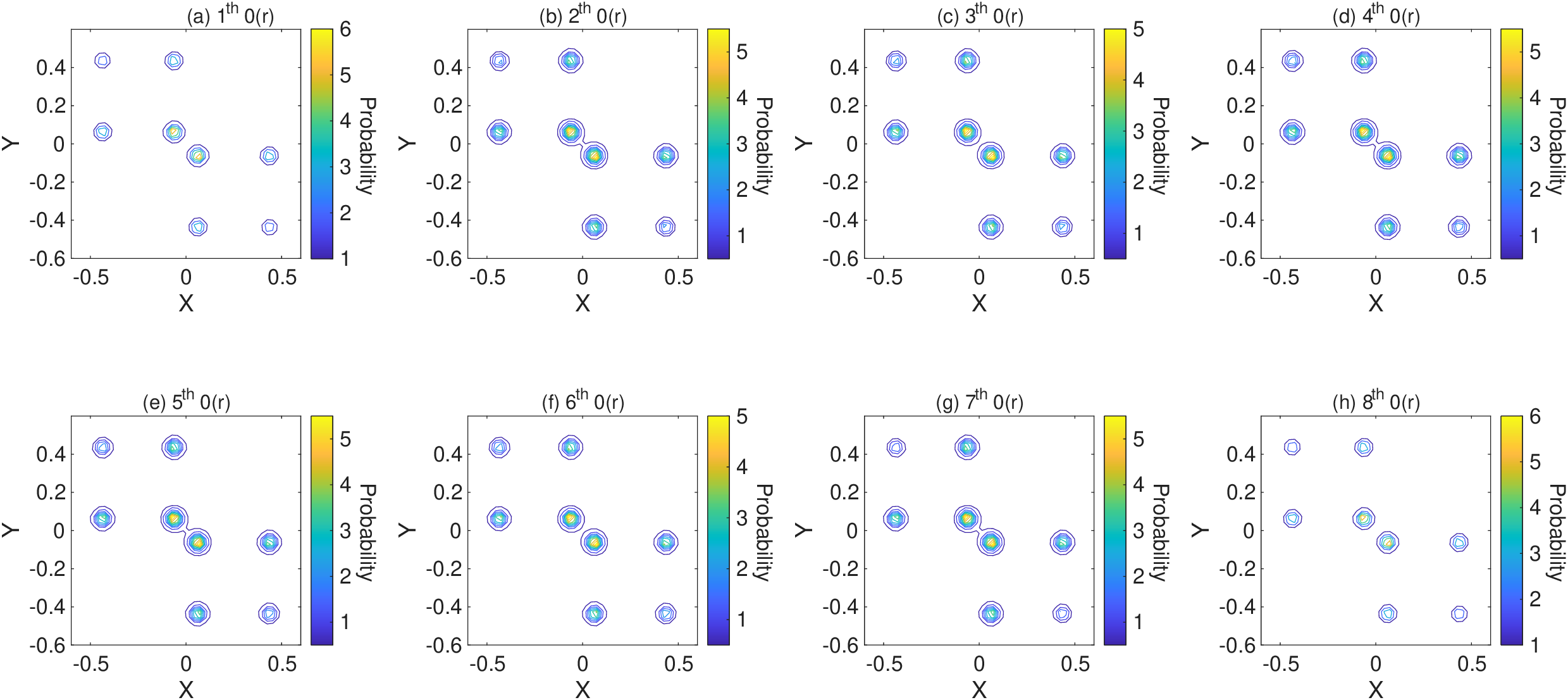}
\caption{{\bfseries Atom-type Wannier functions for the obstructed atomic insulator with opposite flux in Figure. S1(c).}}
\end{figure}

\begin{figure}[h]
\centering
\includegraphics[scale=0.28]{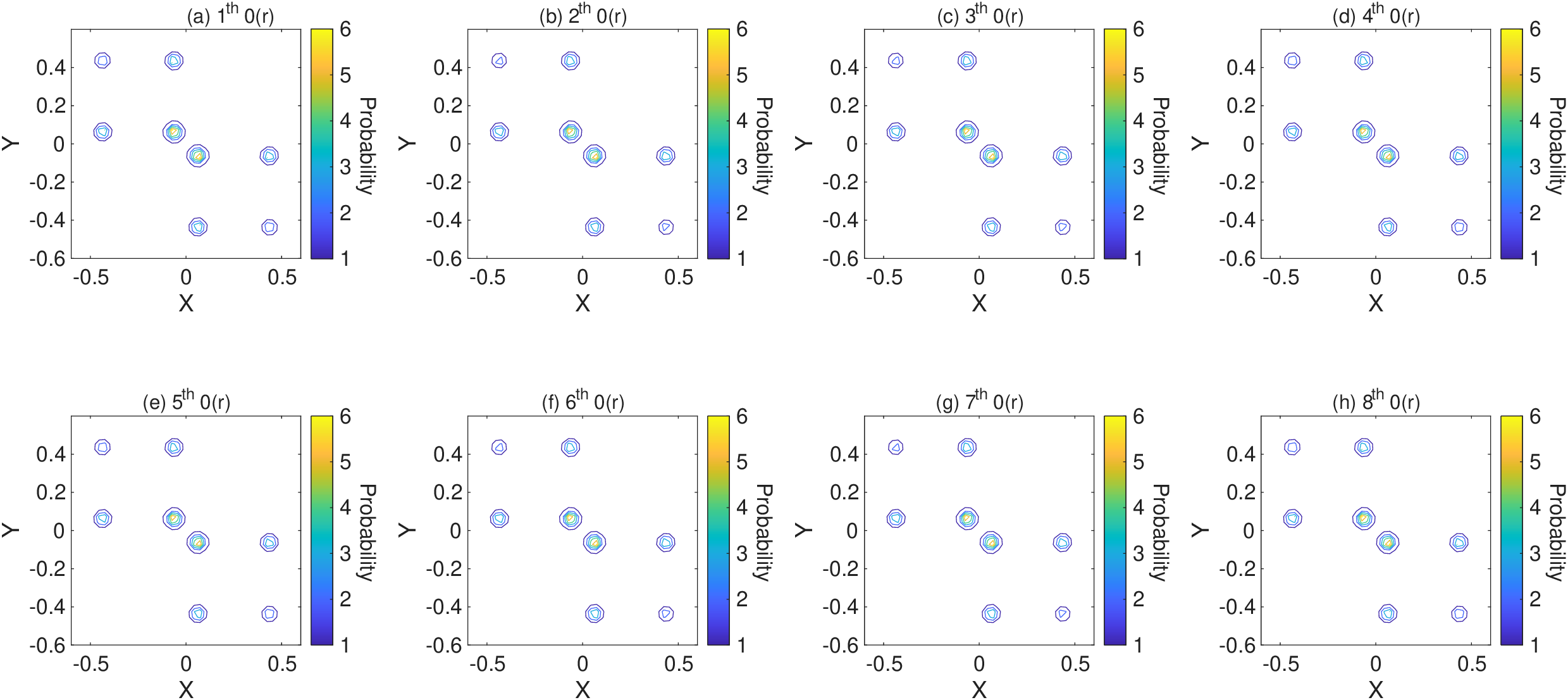}
\caption{{\bfseries Atom-type Wannier functions for the obstructed atomic insulator with the same flux in Figure. S1(e). }}
\end{figure}

\begin{figure}[h]
\centering
\includegraphics[scale=0.28]{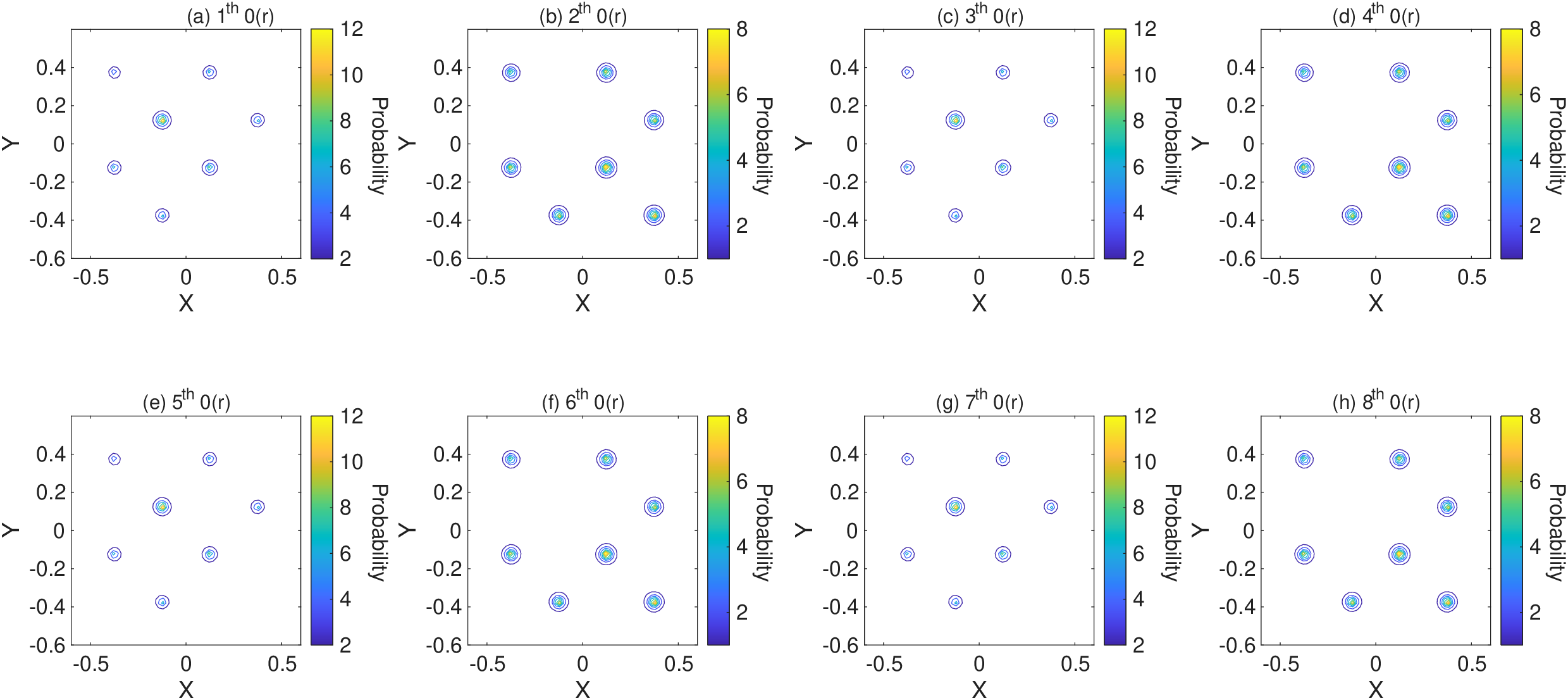}
\caption{{\bfseries Bond-type Wannier functions for the atomic-limit phase in Figure. S1(a). }}
\end{figure}

\begin{figure}[h]
\centering
\includegraphics[scale=0.28]{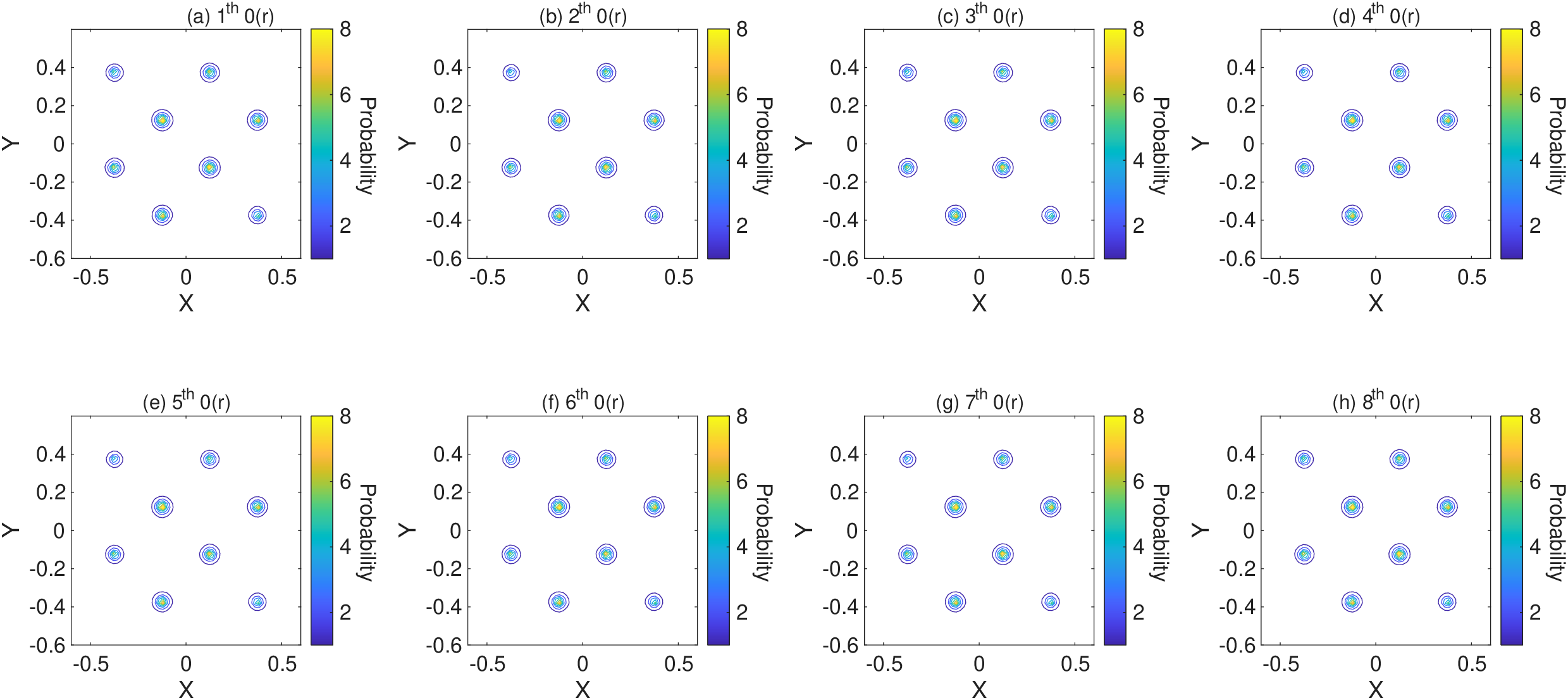}
\caption{{\bfseries Bond-type Wannier functions for the metallic phase in Figure. S1(b). }}
\end{figure}

\begin{figure}[h]
\centering
\includegraphics[scale=0.28]{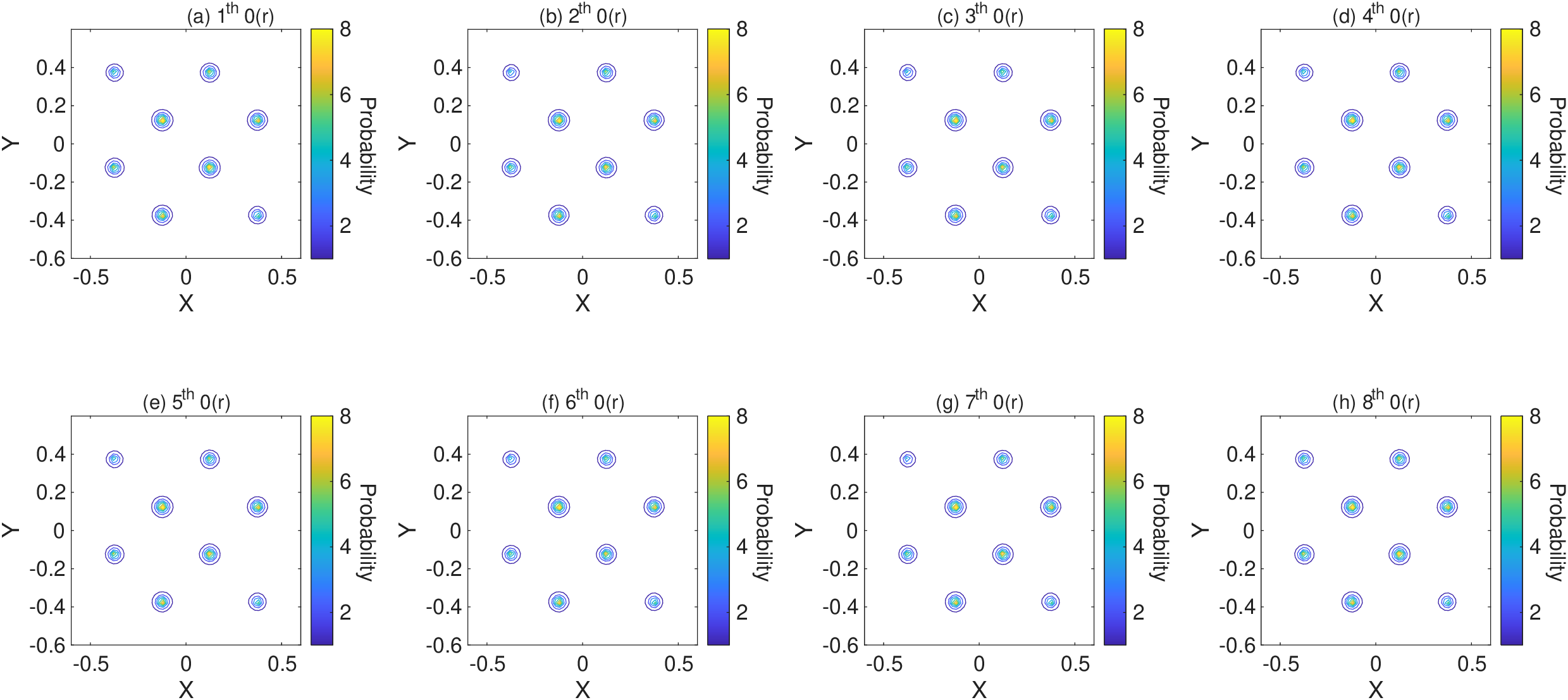}
\caption{{\bfseries Bond-type Wannier functions for the obstructed atomic insulator with opposite flux in Figure. S1(c). }}
\end{figure}

\begin{figure}[h]
\centering
\includegraphics[scale=0.28]{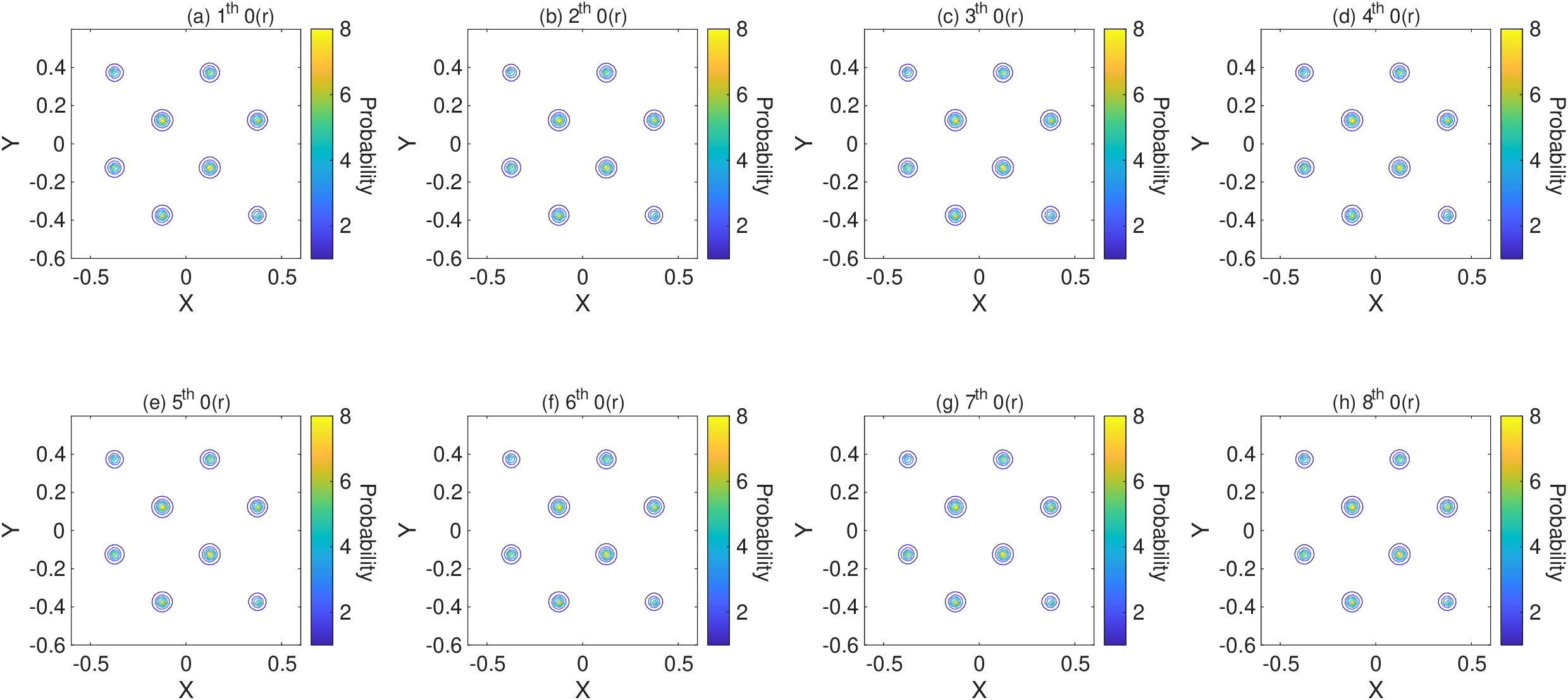}
\caption{{\bfseries Bond-type Wannier functions for the obstructed atomic insulator with the same flux in Figure. S1(e). }}
\end{figure}

\begin{figure}[h]
\centering
\includegraphics[scale=1]{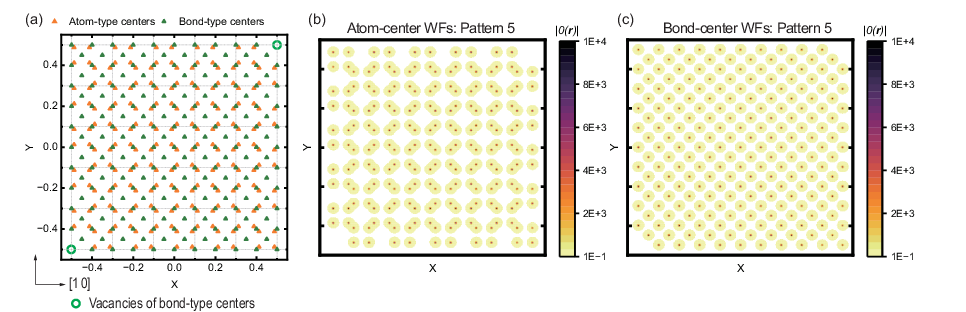}
\caption{{\bfseries Wannier functions for pattern 5.} (a) Positions of atom-type centers (orange triangular) and bond-type centers (green triangular). The bond-center centers with the probability of zero inside the square are labeled by green circle. (b) Wannier functions with atom-type centers for pattern 5. (c) Wannier functions with bond-type centers for pattern 5. }
\end{figure}

\begin{figure}[h]
\includegraphics[scale=1]{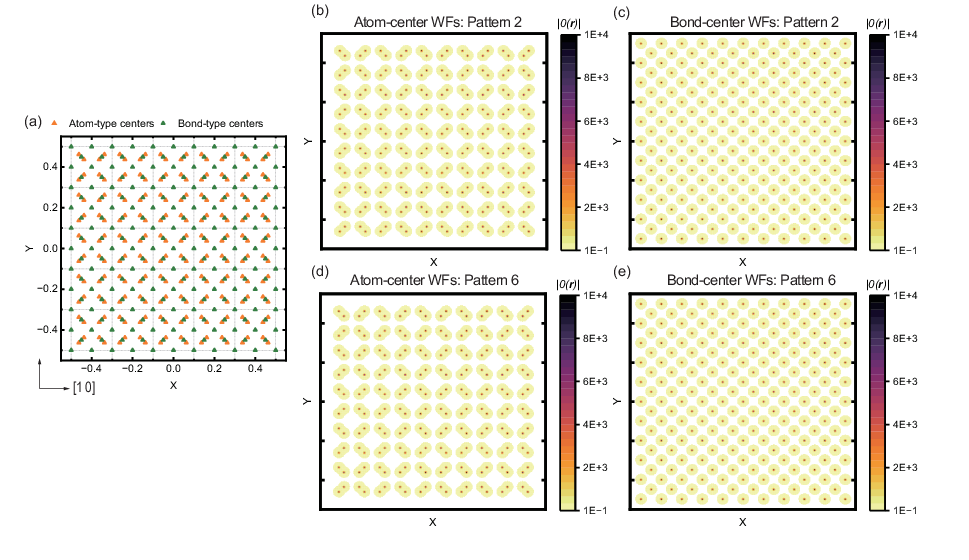}
\caption{{\bfseries Wannier functions for pattern 2 and pattern 6.} (a) Positions of atom-type centers (orange triangular) and bond-type centers (green triangular). (b) Wannier functions with atom-type centers for pattern 2. (c) Wannier functions with bond-type centers for pattern 2. (d) Wannier functions with atom-type centers for pattern 6. (e) Wannier functions with bond-type centers for pattern 6. }
\end{figure}

\begin{figure}[h]
\includegraphics[scale=1]{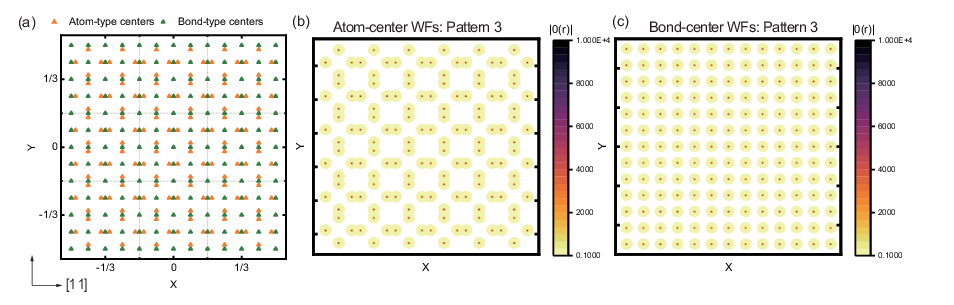}
\caption{{\bfseries Wannier functions for pattern 3.} (a) Positions of atom-type centers (orange triangular) and bond-type centers (green triangular). (b) Wannier functions with atom-type centers for pattern 3. (c) Wannier functions with bond-type centers for pattern 3. }
\end{figure}

\begin{figure}[h]
\includegraphics[scale=1]{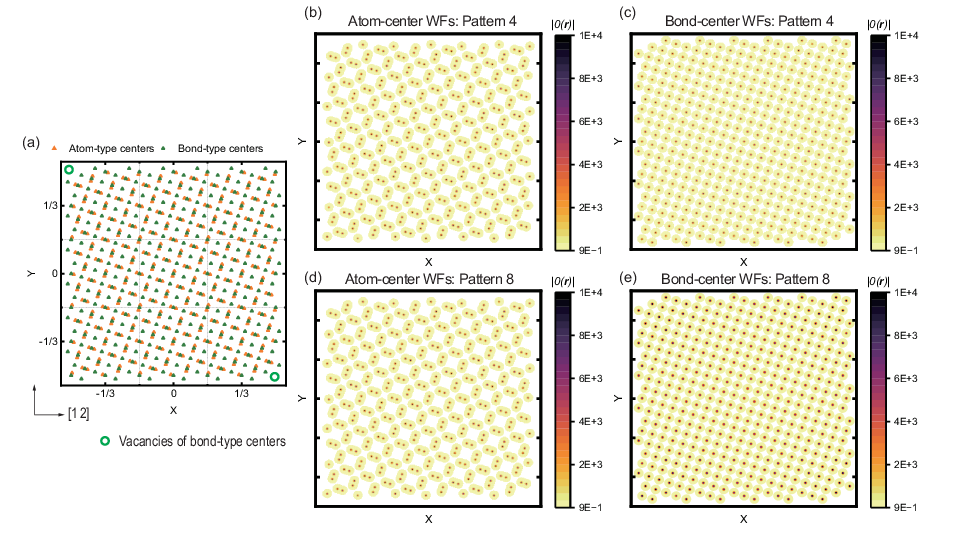}
\caption{{\bfseries Wannier functions for pattern 4 and pattern 8.} (a) Positions of atom-type centers (orange triangular) and bond-type centers (green triangular). The bond-type centers with the probability of zero inside the square are labeled by green circle. (b) Wannier functions with atom-type centers for pattern 4. (c) Wannier functions with bond-type centers for pattern 4. (d) Wannier functions with atom-type centers for pattern 8. (e) Wannier functions with bond-type centers for pattern 8. }
\end{figure}

\clearpage
\newpage

\section{Transformation between TB-model and Dirac fermionic Lagrangian}
The phase transition from the Dirac metal to the OAI by tuning the NN hopping, $\tilde{t}$ = ($\theta + \frac{\pi}{2}$), can also be treated as a $C_2(\mathcal{T})$-symmetric domain wall. In other words, the corner modes can be related to the sharply-changed fermionic modes at the corners and the evolution of mass vector in the Wen-Zee-like term \cite{Wen1992,May2022}. 

The Hamiltonian in Eq.S1 can be transformed into reciprocal space with respect to the 8-component basis: $C_2^{\rm Rotation\,center = 4c}\{ 1, 2, 3, 4\}$ = $\{ 5, 6, 7, 8\}$, $C_4^{\rm Rotation\,center = 2b}\{ 1, 2, 3, 4\}$ = $\{ 2, 3, 4, 1\}$ and $C_4^{\rm Rotation\,center = 2b}\{ 5, 6, 7, 8\}$ = $\{ 6, 7, 8, 5\}$ with $m$ = $\frac{\tilde{t}}{\pi}$ = ($\frac{\theta}{\pi} + \frac{1}{2}$):
\begin{equation}
\begin{aligned}
H^{v1} &= \begin{bmatrix} 
0&\frac{\sqrt{2}}{4}e^{i\phi}&-\frac{1}{4}&\frac{\sqrt{2}}{4}e^{-i\phi}&m&0&0&0\\
\frac{\sqrt{2}}{4}e^{-i\phi}&0&\frac{\sqrt{2}}{4}e^{i\phi}&-\frac{1}{4}&0&me^{ik_x}&0&0 \\
-\frac{1}{4}&\frac{\sqrt{2}}{4}e^{-i\phi}&0&\frac{\sqrt{2}}{4}e^{i\phi}&0&0&me^{i(k_x-k_y)}&0\\
\frac{\sqrt{2}}{4}e^{i\phi}&-\frac{1}{4}& \frac{\sqrt{2}}{4}e^{-i\phi}&0&0&0&0&me^{-ik_y}\\
m&0&0&0&0&\frac{\sqrt{2}}{4}e^{-i\phi}&-\frac{1}{4}&\frac{\sqrt{2}}{4}e^{i\phi}\\
0&me^{-ik_x}&0&0&\frac{\sqrt{2}}{4}e^{i\phi}&0&\frac{\sqrt{2}}{4}e^{-i\phi}&-\frac{1}{4}\\
0&0&me^{i(k_y-k_x)}&0&-\frac{1}{4}&\frac{\sqrt{2}}{4}e^{i\phi}&0&\frac{\sqrt{2}}{4}e^{-i\phi}\\
0&0&0&me^{ik_y}&\frac{\sqrt{2}}{4}e^{-i\phi}&-\frac{1}{4}&\frac{\sqrt{2}}{4}e^{i\phi}&0\\
\end{bmatrix}\\
&=\begin{pmatrix} 
H_{11}^{v1}&H_{12}^{v1} \\
H_{21}^{v1}&H_{22}^{v1}
\end{pmatrix},
\end{aligned}
\end{equation}
where 
\begin{equation}
\begin{aligned}
&H_{11} ^{v1}= \begin{bmatrix}
0&-\frac{1}{4}+\frac{i}{4}&-\frac{1}{4}&-\frac{1}{4}-\frac{i}{4}\\
-\frac{1}{4}-\frac{i}{4}&0&-\frac{1}{4}+\frac{i}{4}&-\frac{1}{4}\\
-\frac{1}{4}&-\frac{1}{4}-\frac{i}{4}&0&-\frac{1}{4}+\frac{i}{4}\\
-\frac{1}{4}+\frac{i}{4}&-\frac{1}{4}&-\frac{1}{4}-\frac{i}{4}&0
\end{bmatrix}\\
&= \frac{-1}{4}(\sigma_0\otimes\sigma_1 + \sigma_1\otimes\sigma_1 + \sigma_1\otimes\sigma_0) + \frac{1}{4} (\sigma_1\otimes\sigma_2 - \sigma_0\otimes\sigma_2) = H_{22}^{v1*} ,\\
&H_{12} ^{v1}= \begin{bmatrix}
m&0&0&0\\
0& m\cos k_x+im\sin k_x&0&0\\
0&0& m\cos (k_x+k_y)+im\sin (k_x+k_y)&0\\
0&0&0& m\cos k_y-im\sin k_y
\end{bmatrix}\\
&= m(\sigma_0 + \sigma_z)\otimes(\sigma_0 + \sigma_z) \\
&+ (m\cos k_x+im\sin k_x) (\sigma_0 + \sigma_z)\otimes(\sigma_0 - \sigma_z)\\
&+ [m\cos (k_x+k_y)+im\sin (k_x+k_y)] (\sigma_0 - \sigma_z)\otimes(\sigma_0 + \sigma_z)\\
&+ (m\cos k_y-im\sin k_y)(\sigma_0 - \sigma_z)\otimes(\sigma_0 - \sigma_z)\\
&= H_{21}^{v1*},\\
\end{aligned}
\end{equation}
so the Bloch Hamiltonian can be written as:
\begin{equation}
\begin{aligned}
&H^{v1} = -\frac{1}{4} \tau_0\otimes(\sigma_0\otimes\sigma_1+\sigma_1\otimes\sigma_1+\sigma_1\otimes\sigma_0)\\
&+ \frac{1}{4} \tau_3\otimes(\sigma_1\otimes\sigma_2 - \sigma_0\otimes\sigma_2)\\
&+m\tau_1\otimes[(\sigma_0 + \sigma_3)\otimes(\sigma_0 +\sigma_3)+\cos k_x(\sigma_0+\sigma_3)\otimes(\sigma_0-\sigma_3) \\
&+\cos (k_x+k_y)(\sigma_0-\sigma_3)\otimes(\sigma_0+\sigma_3)+\cos k_y(\sigma_0-\sigma_3)\otimes(\sigma_0-\sigma_3)] \\
&+m\tau_2\otimes[\sin k_x(\sigma_0 +\sigma_3)\otimes(\sigma_0-\sigma_3)\\
&+\sin (k_x + k_y)(\sigma_0 -\sigma_3)\otimes(\sigma_0+\sigma_3)\\
&-\sin k_y(\sigma_0 -\sigma_3)\otimes(\sigma_0-\sigma_3)].
\end{aligned}
\end{equation}
The model preserves the $C_2\mathcal{T}$ and $C_4$ rotation:
\begin{equation}
\begin{aligned}
\hat{U} &= diag(e^{i\pi/2}, e^{-i\pi/2}\mathcal{K})\otimes diag(e^{i3\pi/4}, e^{i\pi/4}, e^{-i\pi/4}, e^{-i3\pi/4})\\
&=diag(e^{i5\pi/4}, e^{i3\pi/4},e^{i\pi/4}, e^{-i\pi/4},e^{i\pi/4}\mathcal{K}, e^{-i\pi/4}\mathcal{K},e^{-i3\pi/4}\mathcal{K}, e^{-i5\pi/4}\mathcal{K})
\end{aligned}
\end{equation}
to make $\hat{U}H(\bm{k})\hat{U}^{-1}=H(R\bm{k})$, where $\mathcal{K}$ is the complex conjugation operator and $R$ is the combined rotation: $C_2^{\rm Rotation\,center = 4c}\{ 1, 2, 3, 4\}$ = $\{ 5, 6, 7, 8\}$, $C_4^{\rm Rotation\,center = 2b}\{ 1, 2, 3, 4\}$ = $\{ 2, 3, 4, 1\}$ and $C_4^{\rm Rotation\,center = 2b}\{ 5, 6, 7, 8\}$ = $\{ 6, 7, 8, 5\}$. 
The continuum Lagrangian for such Dirac fermion can be written as:
\begin{equation}
\mathcal{L}^{v1} = \overline{\bm{\Psi}}[g^{v1}_0i\partial_t + g^{v1}_1i\partial_x +g^{v1}_2i\partial_y + \bm{m}\cdot g^{v1}_m]\bm{\Psi}.
\end{equation}
where ${\bm{\Psi}}$ and $\overline{\bm{\Psi}}$ = ${\bm{\Psi}}^{\dagger}(\tau_2\mathcal{K})\otimes\sigma^0\otimes\sigma^3$ are 8-component spinors. $\sigma^0\otimes\sigma^3$ is the isospin of the natural Dirac fermion and the $-i$ before $\tau_2\mathcal{K}$ (manual time reversal matrix for spin-1/2 particle: $\mathcal{T}$ = $-i\tau_2\mathcal{K}$) is canceled by the $i$ after $g^{0}$, then we get the full coefficients and find that the defined mass terms are coupled with the reciprocal-spatial variables where the matrix configurations of the coefficients are: 
\begin{equation}
\begin{aligned}
&g^{v1}_0 = (\tau_2\mathcal{K})\otimes\sigma^0\otimes\sigma^3 ,\\
&g^{v1}_1 = m\tau_2\otimes(\sigma_0 + \sigma_3)\otimes(\sigma_0 - \sigma_3), \\
&g^{v1}_2 =m\tau_2\otimes(\sigma_0 - \sigma_3)\otimes(\sigma_0 - \sigma_3),\\
&g^{v1}_m = m\tau_1\otimes(\sigma_0 + \sigma_3)\otimes(-\sigma_0 - \sigma_3).\\
\end{aligned} 
\end{equation}

If we decouple the mass term and reciprocal-spatial variables, the Hamiltonian will be transformed into a shifted version:
\begin{equation}
\begin{aligned}
H^{v2} &= \begin{bmatrix} 
0&\frac{\sqrt{2}}{4}e^{i\phi}&-\frac{1}{4}&\frac{\sqrt{2}}{4}e^{-i\phi}&m&0&0&0\\
\frac{\sqrt{2}}{4}e^{-i\phi}&0&\frac{\sqrt{2}}{4}e^{i\phi}&-\frac{1}{4}&0&m&0&0\\
-\frac{1}{4}&\frac{\sqrt{2}}{4}e^{-i\phi}&0&\frac{\sqrt{2}}{4}e^{i\phi}&0&0&m&0\\
\frac{\sqrt{2}}{4}e^{i\phi}&-\frac{1}{4}&\frac{\sqrt{2}}{4}e^{-i\phi}&0&0&0&0&m\\
m&0&0&0&0&\frac{\sqrt{2}}{4}e^{-i\phi-ik_x}&-\frac{1}{4}e^{ik_y-ik_x}&\frac{\sqrt{2}}{4}e^{i\phi+ik_y}\\
0&m&0&0&\frac{\sqrt{2}}{4}e^{i\phi+ik_x}&0&\frac{\sqrt{2}}{4}e^{-i\phi+ik_y}&-\frac{1}{4}e^{ik_x+ik_y}\\
0&0&m&0&-\frac{1}{4}e^{ik_x-ik_y}&\frac{\sqrt{2}}{4}e^{i\phi-ik_y}&0&\frac{\sqrt{2}}{4}e^{-i\phi+ik_x}\\
0&0&0&m&\frac{\sqrt{2}}{4}e^{-i\phi-ik_y}&-\frac{1}{4}e^{-ik_x-ik_y}&\frac{\sqrt{2}}{4}e^{i\phi-ik_x}&0\\
\end{bmatrix}\\
&=\begin{pmatrix} 
H_{11}^{v2}&H_{12}^{v2} \\
H_{21}^{v2}&H_{22}^{v2}
\end{pmatrix},
\end{aligned}
\end{equation} 
so 
\begin{equation}
\begin{aligned}
&H^{v2} =(\tau_0 + \tau_3)\otimes[ \frac{-1}{4}(\sigma_0\otimes\sigma_1 + \sigma_1\otimes\sigma_1 + \sigma_1\otimes\sigma_0) + \frac{1}{4} (\sigma_1\otimes\sigma_2 - \sigma_0\otimes\sigma_2) ] + m\tau_1\otimes\sigma_0\otimes\sigma_0\\
&+(\tau_0 - \tau_3)\otimes[ \\
&-\frac{1}{4}(\cos k_x+\sin k_x) (\sigma_0+\sigma_3)\otimes\sigma_1\\
&+\frac{1}{4}(\cos k_x-\sin k_x)(\sigma_0+\sigma_3)\otimes\sigma_2\\
&-\frac{1}{4}\cos (k_x-k_y) \sigma_1\otimes(\sigma_0+\sigma_3)\\
&- \frac{1}{4}\sin (k_x-k_y)\sigma_2\otimes(\sigma_0+\sigma_3)\\
&-\frac{1}{4}(\cos k_y+\sin k_y)(\sigma_+\otimes\sigma_+ + \sigma_-\otimes\sigma_-) \\
&-\frac{1}{4}(\cos k_y+\sin k_y)\sigma_1\otimes\sigma_2\\
&-\frac{1}{4}(\cos k_y-\sin k_y) (\sigma_+\otimes\sigma_- + \sigma_-\otimes\sigma_+) \\
&- \frac{1}{4}\cos (k_x+k_y)\sigma_1\otimes(\sigma_0-\sigma_3)\\
&+\frac{1}{4}\sin (k_x+k_y) \sigma_2\otimes(\sigma_0-\sigma_3)\\
&-\frac{1}{4}(\cos k_x-\sin k_x)(\sigma_0-\sigma_3)\otimes\sigma_1\\
&+\frac{1}{4}(\cos k_x+\sin k_x)(\sigma_0-\sigma_3)\otimes\sigma_2]. \\
\end{aligned}
\end{equation}
The continuum Lagrangian for such Dirac fermion can be written as:
\begin{equation}
\mathcal{L}^{v2} = \overline{\bm{\Psi}}[g^{v2}_0i\partial_t + g^{v2}_1i\partial_x +g^{v2}_2i\partial_y + \bm{m}\cdot g^{v2}_m]\bm{\Psi}.
\end{equation}
 where the matrix configurations of the coefficients become:
\begin{equation}
\begin{aligned}
&g^{v2}_0 = (\tau_2\mathcal{K})\otimes\sigma^0\otimes\sigma^3 ,\\
&g^{v2}_1 = (\tau_0 - \tau_3)\otimes[(\sigma_0 + \sigma_3)\otimes \sigma_2 - (\sigma_0 + \sigma_3)\otimes \sigma_1-(\sigma_0 - \sigma_3)\otimes \sigma_2 + (\sigma_0 - \sigma_3)\otimes \sigma_1],\\
&g^{v2}_2= (\tau_0 - \tau_3)\otimes[\sigma_+\otimes\sigma_+ - \sigma_-\otimes\sigma_- - \sigma_1\otimes\sigma_1 + \sigma_+\otimes\sigma_- - \sigma_-\otimes\sigma_+],\\
&g^{v2}_m =-m\tau_1\otimes\sigma_0\otimes\sigma_3.
\end{aligned} 
\end{equation}

From the first case of transformation where the mass of Dirac fermion is coupled with reciprocal-spatial variables, we write the wave functions of four edges as: $\Psi_{+x}$, $\Psi_{-x}$, $\Psi_{+y}$, $\Psi_{-y}$, then the gapped one-dimensional domain wall Fermion is
\begin{equation}
\begin{aligned}
\mathcal{L}_{\rm edge\,mass} = M\Psi_{+x}^{\dagger}\Psi_{+x} + M\Psi_{+y}^{\dagger}\Psi_{+y}- M\Psi_{-x}^{\dagger}\Psi_{-x} - M\Psi_{-y}^{\dagger}\Psi_{-y} 
\end{aligned}
\end{equation}
under $C_2$ rotation. In this case, Jackiw-Rebbi zero mode appears at the corner which connect two adjacent edges with opposite mass and has half-integer charge. 

Then the Dirac fermionic field with Wen-Zee-like topological response:
\begin{equation}
\mathcal{L}_{\rm WZ} = \frac{{\rm sgn(m)}-1}{2\pi}\epsilon^{\mu\nu\rho}\omega_\mu \partial_\nu A_\rho
\end{equation}
under the covariant derivative: $D_\mu = \partial_\mu -iA_\mu -i\omega_\mu S$ where $S$ is the rotation matrix of the Dirac fermion in the scenario of the $C_2$ confinement.
The existence of Dirac fermion mode is further proved by the lifted degeneracy under $C_2$-breaking perturbation as shown in Fig. S17. 

If we cleave the OAI by leaving four corners all with NN dangling bonds, $i.e.$, containing the mass of Dirac fermions, we observe the four-fold degenerate zero mode in a 4.5$\times$4.5 superlattice. Hence, whether the mass terms in the models appear at the corner regions determines the emergence of corner states by comparing with the $C_4$-symmetric pattern 3 and pattern 7. 

\begin{figure}[h]
\centering
\includegraphics[scale=1]{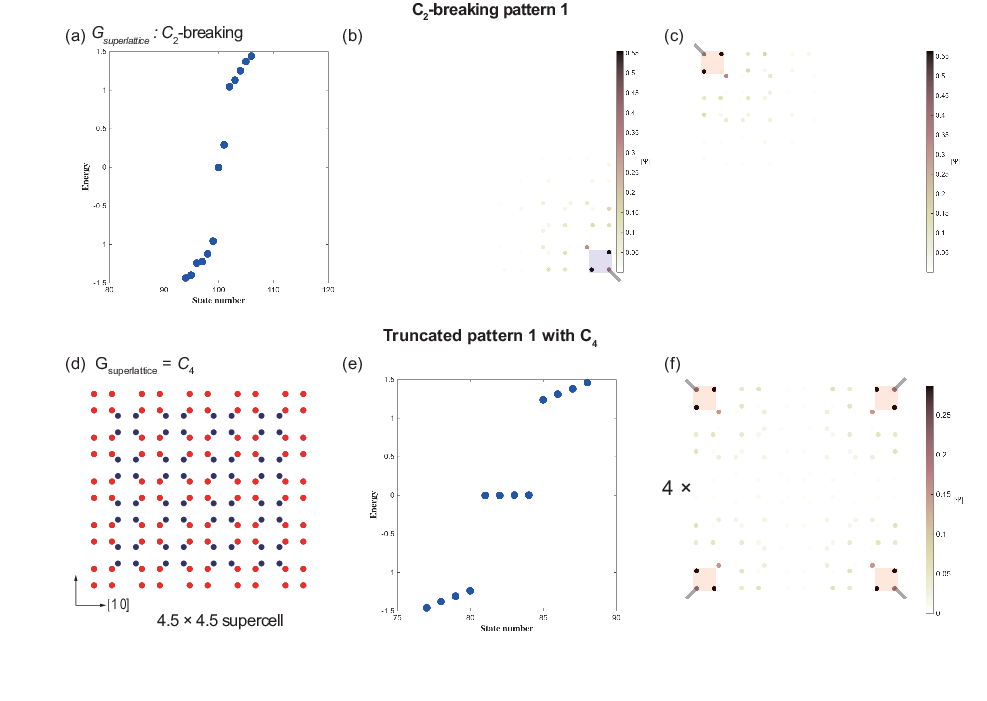}
\vspace{-1.2 cm}
\caption{{\bfseries Evidence of Dirac fermion mode rather than Majorana zero mode}: (a-c) Lifted zero-energy degeneracy in pattern 1 with $C_2$-breaking perturbation in which (a) is the energy spectrum and (b), (c) are the corresponding real-space wavefunctions. {\bfseries Mass term as the reason for the emergence of zero-energy corner modes}: (d-f) Four-fold zero-energy degeneracy in 4.5$\times$4.5 truncated pattern 1 with $C_4$ symmetry. (d) is the lattice configuration. (e) is the energy spectrum. (f) is the real-space wavefunction of the four-fold degenerate zero modes. The two sub-lattices with NN dangling bonds at WP 4$c$ are labeled by the translucent red and blue squares with gray short lines. }
\end{figure}

\clearpage
\newpage
\section{Valence-bond-(Dirac-mass-)configured lattice gauge }

Although we have known that the NN dangling bonds (as the Dirac mass which induces the zero-energy gap) in the corner regions determine the emergence of zero modes, we do not know why the NN dangling bonds along $\pm x$, $\pm y$ in patterns 3 and 7 suppress the zero modes. It implies that the local rotation of the NN dangling bonds also impacts the zero modes, namely, the rotation symmetry in $G_{\rm WP=4c}$. By this consideration, we investigate the lattice gauge around the Fermi level, namely, "the real-space valence-bond configuration". To get such a valence bond-dependent lattice gauge, we perform coarse graining for the original Hamiltonian [see Figs. S18(a-d)]:
\begin{itemize}
\item Treat the sub-lattices distinguished by the mass-relative hopping as spinful orbital degrees, labeled by the Pauli matrices $\tau_i$, $i= x, y, z$.
\item Keep the hoppings related to the Dirac mass and neglect all other hoppings. Classify all the mass-relative electron transfer as intra-orbital hoppings and the mass-relative inter-sublattice electron transfer as inter-orbital hoppings. 
\item We find that the intra-orbital mass-dependent hoppings preserve the original real-space local symmetry which can be originated from the $G_{\rm WP = 4c}$; While the inter-orbital mass-dependent hoppings do not obey $G_{\rm WP = 4c}$ and carry an inter-level flux. 
\end{itemize}
Hence, for the Hamiltonian in Eq. (1), we get:
\begin{equation}
\begin{aligned}
&H^{\rm C_2 \,coarse\,grain} = \\
&\begin{bmatrix} (e^{-ik_x}+e^{ik_x})+2(e^{-ik_y}+e^{ik_y}) + 3(e^{ik_x-ik_y} + e^{ik_y-ik_x}) ; (e^{-ik_x})+ (e^{-ik_y}) + (e^{ik_x-ik_y})\\
(e^{ik_x})+ (e^{ik_y}) + J(e^{ik_y-ik_x}); (e^{-ik_x} + e^{ik_x})+ 2(e^{-ik_y} + e^{ik_y}) + 3(e^{ik_x-ik_y} + e^{ik_y-ik_x}) 
\end{bmatrix}\\
&\equiv\ d_x \tau_x + d_y \tau_y + d_z \tau_z,
\end{aligned}
\end{equation}
where the remaining intra-orbital mass-relative hoppings preserve the $C_2$ symmetry, which is from the real-space local symmetry $G_{\rm WP = 4c}$, while the remaining inter-orbital mass-relative hoppings serve as the off-diagonal matrix elements.  

\begin{figure}[h]
\centering
\includegraphics[scale=1]{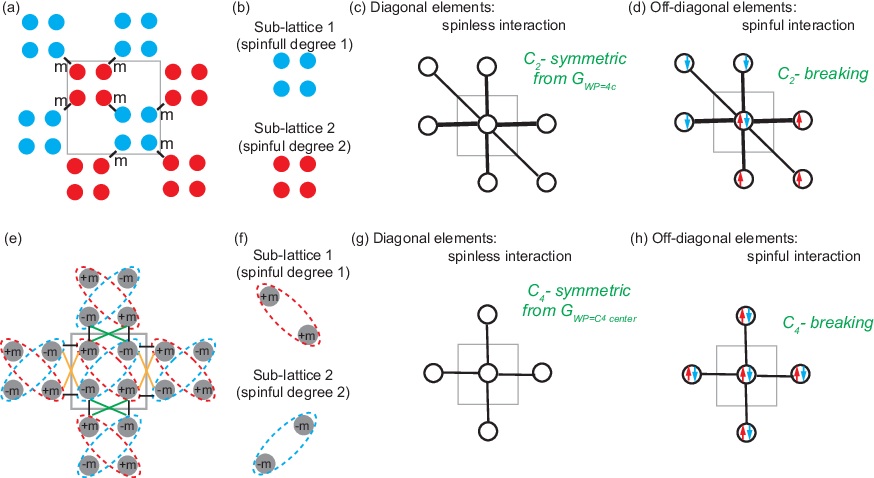}
\caption{{\bfseries "Coarse graining“ of Hamiltonians for extracting valence-bond-(Dirac-mass-)dependent lattice gauge.} (a) Original hopping diagram for the Hamiltonian in Eq. (1). (b) Two sub-lattices as the two super-sites in the model of Eq. (1). (c) Hopping diagram for the mass-dependent intra-orbital interactions that obey $C_2$ (from $G_{\rm WP = 4c}$) rotation symmetry. (d) Hopping diagram for the mass-dependent inter-orbital interactions that do not obey $C_2$ but carry a flux. (e) Original hopping for the Hamiltonian of a HOTI with second-order corner states at four corners \cite{May2022}. (f) Two sites with different signs of mass-dependent on-site terms as the two super-sites. (g) Hopping diagram for the mass-dependent intra-orbital interactions which obeys $C_4$ rotation symmetry. (h) Hopping diagram for the mass-dependent inter-orbital interactions which do not obey $C_4$ but carry a flux.}
\end{figure}

To highlight the anisotropy of the HOTI in this work, we perform the same operation in previous HOTI with second-order corner states at four corners starting from the original Hamiltonian \cite{May2022}:
\begin{equation}
H^{\rm C_4\,HOTI}= \sin k_x \sigma_3\otimes \sigma_2 - \sin k_y\sigma_3\otimes \sigma_1 + (\cos k_x - \cos k_y) \sigma_2\otimes \sigma_0 + (m + \cos k_x + \cos k_y)\sigma_3\otimes \sigma_3
\end{equation}
to a simplified model with two spinful orbital degrees determined by the sign of the on-site Dirac-mass hopping term  [see Figs. S18(e-h)]:
\begin{equation}
\begin{aligned}
&H^{\rm C_4\,coarse\,grain} = \\
&\begin{bmatrix} (e^{-ik_x}+e^{ik_x})+(e^{-ik_y}+e^{ik_y}) &\sin k_y + i\sin k_x\\
\sin k_y - i\sin k_x &(e^{-ik_x} + e^{ik_x})+ (e^{-ik_y} + e^{ik_y})
\end{bmatrix}\\
&\equiv\ d_x \tau_x + d_y \tau_y + d_z \tau_z.
\end{aligned}
\end{equation}
where the remaining intra-orbital mass-relative hoppings preserve the $C_4$ symmetry, which is from the real-space $C_4$-rotation center. 

Notably, regardless of whether they are the "inter-site hoppings" or the "on-site energies with different sign", the critical mass terms drive a real-space critical electron transfer that induces the zero-energy phase transition from metallic states to insulating states. This relative orientation between crystal cleavage and bonding is similar to the nesting relation in momentum space \cite{Wieder2018,Hwang2019,Ahn2019,Wieder2020}. 

\clearpage
\newpage
\section{Formulas of entanglement entropy and density-weighted energy}

Entanglement entropy in a tight-binding model measures the quantum entanglement between two spatial regions. It is obtained from the correlation matrix $C_{ij}$ which is defined by
\begin{equation}
C_{ij} = \sum_{n=1}^{N_{\rm occupied}}a_{n}^{\alpha,\bm{r}_i*} a_{n}^{\alpha,\bm{r}_j}.
\end{equation}
$C_{ij}$ is complex and possesses eigenvalues of $\{\lambda_1, \lambda_2,... \}$. Then the entropy is defined by
\begin{equation}
\label{entropy}
S(\rho) = -\sum_{j=1}^{N_{\rm sub\, system}}[\lambda_j ln(\lambda_j) + (1-\lambda_j)ln(1-\lambda_j)].
\end{equation}
The mathematical expression and the monotonicity of this integrant are shown Fig. S20. It turns out that the lower entropy at a site represents that the components of all occupied states projected on the site are less.

\begin{figure}[h]
\centering
\includegraphics[scale=1.4]{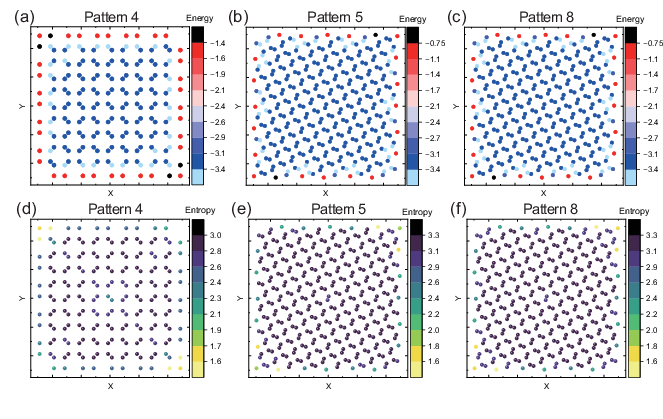}
\caption{{\bfseries Real-space distribution of the density-weighted energy and entanglement entropy in the topologically nontrivial patterns.} (a, b, c) Real-space distribution of density-weighted energy of occupied states for patterns 4, 5, 8. (d, e, f) Real-space distribution of the one-site subsystem entanglement entropy of occupied states for patterns 4, 5, 8.}
\end{figure}

\begin{figure}[h]
\centering
\includegraphics[scale=0.5]{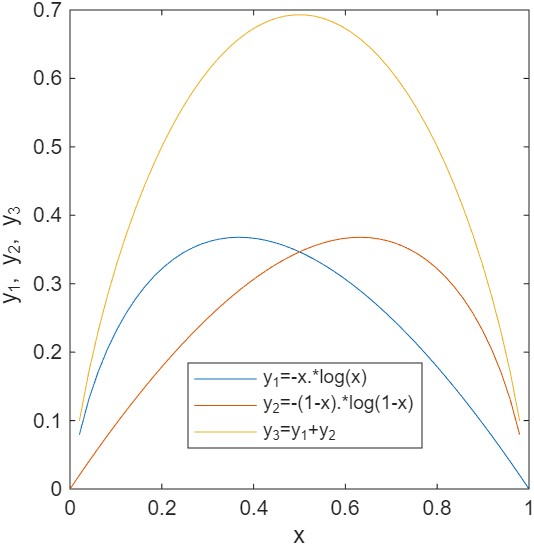}
\caption{{\bfseries Mathematical expression and the monotonicity} of the integrant $-[\lambda_j ln(\lambda_j) + (1-\lambda_j)ln(1-\lambda_j)]$ in Eq.\ref{entropy}: $y_3 = y_1 + y_2 = [-x*log(x)] + [-(1-x)*log(1-x)]$.}
\end{figure}
The real-space distribution of density-weighted energy can be calculated in the same way with corner charge in Eq.\ref{cornercharge}:
\begin{equation}
E(\bm{r}) = \sum_{n=1}^{N_{\rm occupied}}E_n\sum_\alpha^{N_{\rm orbital}} |a_n^{\alpha,\bm{r}}|^2,
\end{equation}
where $E_n$ is the energy of $n^{th}$ energy level.

\end{document}